\documentclass[journal,twocolumn,10pt,twoside]{IEEEtran}
\usepackage{graphicx,cite,epsfig,amssymb,amsmath,amsfonts,bbm}
\usepackage{pifont}
\usepackage{stmaryrd}
\usepackage{bbding}
\usepackage{subfigure}
\usepackage{setspace}
\usepackage{color}
\usepackage{slashbox}
\usepackage{algorithm,algorithmic}
\usepackage{bm}
\usepackage{subfigure}
\usepackage{booktabs}
\usepackage{enumerate}
\usepackage{multirow}
\usepackage{framed}
\newtheorem{lemma}{Lemma}
\newtheorem{theorem}{Theorem}
\allowdisplaybreaks[4]

\begin{document}

\title{Exploiting Matrix Information Geometry for Integrated Decoding of Massive Uncoupled Unsourced Random Access}
\author{\normalsize
Feiyan Tian, Xiaoming Chen, Chongwen Huang, and Zhaoyang Zhang

\thanks{Feiyan Tian, Xiaoming Chen, Chongwen Huang, and Zhaoyang Zhang are with the College of Information Science and Electronic Engineering, Zhejiang University, Hangzhou 310027, China (e-mail: \{tian\_feiyan, chen\_xiaoming, chongwenhuang, ning\_ming\}@zju.edu.cn).}
}\maketitle

\begin{abstract}
In this paper, we explore an efficient uncoupled unsourced random access (UURA) scheme for 6G massive communication.
UURA is a typical framework of unsourced random access that addresses the problems of codeword detection and message stitching, without the use of check bits.
Firstly, we establish a framework for UURA, allowing for immediate decoding of sub-messages upon arrival.
Thus, the processing delay is effectively reduced due to the decreasing waiting time.
Next, we propose an integrated decoding algorithm for sub-messages by leveraging matrix information geometry (MIG) theory. Specifically, MIG is applied to measure the feature similarities of codewords belonging to the same user equipment, and thus sub-message can be stitched once it is received. This enables the timely recovery of a portion of the original message by simultaneously detecting and stitching codewords within the current sub-slot.
Furthermore, we analyze the performance of the proposed integrated decoding-based UURA scheme in terms of computational complexity and convergence rate. Finally, we present extensive simulation results to validate the effectiveness of the proposed scheme in 6G wireless networks.
\end{abstract}


\section{Introduction}
Massive communication has been identified as one of usage scenarios of 6G wireless networks \cite{6G,5G1,5G2}. It aims to satisfy the requirements of massive connectivity, small payload, and low latency for user equipments (UEs) in the Internet of Things (IoT). This necessitates efficient uplink random access solutions.
In cellular IoT networks, although there is a large number of UEs, only a small set of them are active within a given interval and needs to transmit data to the base station (BS). In this case, traditional grant-based random access schemes may not be suitable due to their high access latency and signaling overhead \cite{GF0}.
Consequently, the introduction of grant-free random access schemes becomes essential to support massive connectivity but with sporadic data traffic \cite{GF1,GF2}.

The first kind of typical grant-free random access protocol is sourced random access (SRA) \cite{GF0,GF1,GF2}. Under this random access protocol, the identities and channel information of active UEs need to be acquired by sending unique pilot sequences before data transmission. However, the use of long pilots often occupies a significant portion of the packet length to ensure accurate information acquisition, resulting in resource wastage.
To address it, the authors of \cite{bliSRA} proposed a blind goal-oriented massive access scheme. The identities of active stationary devices were indirectly obtained by exploiting angular information of channels, effectively saving the resource.
On the other hand, a new kind of grant-free random access protocol called unsourced random access (URA) has emerged \cite{UN1}. Herein, the BS is only interested in the transmitted messages and not concerned about the identities or channel states of active UEs \cite{UN2,UN3,UN4}. Without the need for pilots used in SRA, URA achieves substantial reductions in wireless resource consumption and end-to-end delay, especially in scenarios with massive connectivity \cite{UN1,UN5}.
In fact, URA is designed for the practical IoT scenarios where a large number of UEs have their codebooks preconfigured during production, such as the massive sensors in smart factories. This allows all UEs to map their messages to codewords using a shared codebook, eliminating the inconvenience of the situation where each device has its own distinct codebook.
To avoid excessive codebook size and unbearable computational complexity of codeword detection, which are determined by the message length, a segmentation strategy is introduced. Each message is fragmented to multiple sub-messages, which are independently sent in segmented transmission sub-slots based on a smaller and fixed codebook \cite{UN6}.
Currently, there are two primary frameworks for segmented URA. One of them is coupled URA (CURA) \cite{UN5}. In this framework, the message fragments of a certain active UE are transmitted across multiple sub-slots. These fragments are coupled by adding redundant check bits to the end of each fragment, facilitating the stitching of fragmented messages at the receiver. Based on this framework, several CURA schemes have been studied from different perspectives \cite{UN8,UN9,UN10}.
For instance, at the sender, the segmentation of original messages and the coupling of fragmented messages can be carried out by exploiting the coding rules of tree code. After the segmented messages are mapped to the codewords using a common codebook and then transmitted, at the BS, Bayesian or non-Bayesian methods can be employed to detect the active codewords in each sub-slot. Subsequently, the corresponding tree decoder connects the detected codewords to recover the original messages based on the redundancy.

In general, the redundant check bits introduced by CURA is limited compared to the pilot sequences used in SRA. Yet, in order to ensure the correctness and uniqueness of concatenation results, the length of additional parity bits added to the tail of message fragment is usually equal to or greater than the length of the fragment itself. This decreases the spectral efficiency of system, which is unacceptable when supporting a large number of UEs with limited wireless resources.
To overcome these challenges, an alternative framework for URA called uncoupled URA (UURA) has been proposed \cite{UN11}. In UURA, redundancy information to couple message fragments is eliminated and codeword stitching is achieved through clustering-based methods rather than checkbits-based ones. To be specific, at the BS, the detected codewords across multiple sub-slots that originate from the same active UE are concatenated according to the correlations of the UE's instantaneous channels across sub-slots.
Based on the above uncoupled framework, the researchers in \cite{UN12} investigated a UURA scheme by utilizing the angular domain sparsity of channels. This scheme demonstrates improved performance and higher spectral efficiency.
Additionally, a modified bilinear vector approximate message passing algorithm was proposed for UURA in \cite{UN13}, where the codeword concatenation was omitted by directly detecting the codeword assignment block matrix of all sub-slots. Furthermore, the authors of \cite{UN14} studied a tensor-based UURA scheme. By leveraging the structure of low-rank tensors and the uniqueness of tensor decomposition, multi-user separation and single-user demapping were well achieved at the receiver.

It is found that the existing UURA schemes may require more processing time for codeword concatenation. Whether the codewords are detected first and then stitched \cite{UN11,UN12} or codeword stitching is directly cancelled \cite{UN13,UN14}, the messages can only be recovered when the complete messages of the entire block length have been received.
Such processing delay may not be suitable for applications with low latency requirements in massive communication scenarios, such as video IoT. Generally, video IoT applications, including video meetings, smart TVs, telecommuting, online classes, and more, demand real-time transmission of multimedia data \cite{VIoT}. In such applications, the encoded data needs to be transmitted immediately at the sender and the received data requires to be timely decoded at the receiver to minimize the waiting time.
In this context, although UURA is a promising access protocol suited for video IoT due to its low end-to-end delay, there is still room for further reduction of processing latency in existing efforts of UURA. Therefore, it is imperative to design a UURA scheme with low processing delay to support massive UEs generating high-volume streaming media transmissions.

Intuitively, in order to decrease the processing delay of UURA schemes, it would be advantageous to detect and concatenate each sub-message in a sub-slot as soon as it is received, rather than waiting for all sub-messages to be transmitted. In other words, the codeword detection and stitching are implemented in an integrated manner rather than two separate parts, which can significantly reduce the waiting time and enable decoding while transmitting. With this aim, we try to design a novel codeword stitching method integrating codeword detection. In general, codeword stitching in UURA can be viewed as a data clustering task. In other words, the feature similarities of codewords belonging to the same active UE should be exploited to stitch the sub-messages. Typically, the types of features used for clustering include low-order statistical features such as mean and variance, as well as high-order statistical features such as covariance matrix.
In the case of a smaller number of available clustering samples (in the existing pattern of separate codeword detection and stitching after receiving all sub-messages, the size of the clustering samples is $K_a\times L$, where $K_a$ represents the number of active UEs and $L$ denotes the total number of sub-slots. While in the proposed pattern of integrated codeword detection and stitching within the current sub-slot, the size of clustering samples is $K_a\times l$ with $l\leq L$ being the index of the current sub-slot. The sample size is limited compared to the former situation), low-order statistical features may be more easily influenced by outliers, leading to unreliable clustering results. Therefore, traditional clustering methods that utilize low-order statistical features for separate codeword stitching may not be suitable.
Instead, an efficient approach is needed to extract and analyze high-order statistical features, which contain more comprehensive data information, enhancing the clustering robustness for timely codeword stitching \cite{class}.
Recently, matrix information geometry (MIG) theory has been proposed to analyze high-order matrix-valued data using the geometric methods \cite{MIG0}. It provides a way to efficiently measure distances and proximities between matrices, making it applicable for exacting feature similarities from codeword covariance matrices of the same active UE in the proposed integrated decoding scenarios.
At present, MIG has been used for radar and maritime target detection to improve the discrimination between targets and clutter \cite{MIG1,MIG2,MIG3}. Furthermore, MIG has demonstrated its potential in data clustering tasks by assigning data points to predefined classes based on their proximity to specific regions in the manifold \cite{MIG4}. Inspired by this, by analyzing the geometric similarities of codeword covariance matrices from the same active UE in multiple sub-slots according to the theory of MIG, we develop a novel codeword concatenation method to facilitate integrated decoding and improve the processing flow of UURA.

In this paper, we aim to provide a latency-efficient UURA scheme for 6G massive communications. The contributions of this paper are as follows:

\begin{enumerate}
\item We propose a UURA framework that enables immediate decoding, in which each sub-message within a sub-slot is detected and stitched simultaneously upon arrival at the receiver. This eliminates the need to wait for the complete message before decoding begins.

\item We design a low-complexity integrated decoding algorithm by leveraging the theory of MIG, which efficiently performs codeword detection and stitching in each sub-slot by utilizing the geometric similarities of codeword covariance matrices.

\item We analyze the computational complexity and convergence rate of the proposed MIG-aided integrated decoding-based UURA scheme. Through analysis, we validate the feasibility and effectiveness of the proposed integrated decoding algorithm.

\end{enumerate}

The rest of this paper is organized as follows: Section II introduces the system model of uncoupled unsourced random access in 6G wireless networks. Section III designs a decoding algorithm integrating codeword detection and stitching for the proposed massive uncoupled unsourced random access framework. Section IV analyzes the computational complexity and convergence rate of the proposed decoding algorithm. Then, simulation results are given in Section V to validate the effectiveness of the proposed algorithm. Finally, Section VI concludes the paper.

\emph{Notations}: Bold upper (lower) letters denote matrices (column vectors), $(\cdot)^T$ denotes transpose, $(\cdot)^H$ denotes conjugate transpose, $\mathbb{C}^{a \times b}$ denotes a complex matrix or vector of dimension $a \times b$, $\mathcal{CN}(\textbf{x},\textbf{Y})$ denotes the complex Gaussian distribution of a vector with mean $\textbf{x}$ and covariance $\textbf{Y}$, $\textmd{P}(\cdot|\cdot)$ denotes the conditional probability, $\nabla$ and $\partial$ denote the gradient and sub-gradient, $\|\cdot\|_2$ and $\|\cdot\|_F$ denote the 2-norm of a vector and Frobenius norm of a matrix respectively, and $\textmd{tr}(\cdot)$ denotes the trace of a matrix.
\section{System Model}
Consider an UURA-based 6G massive communication system comprising one BS equipped with $M$ antennas and $K_{\textmd{tot}}$ single-antenna IoT UEs. All UEs are served by the BS over the same time-frequency resource block. In massive communication systems, there are usually a large number of potential UEs but only a small set of $K_a$ UEs, denoted by $\mathcal{K}_a$, are active, which have burst data requests in a certain time slot, i.e., $K_a \ll K_{\textrm{tot}}$. It is assumed that $K_a$ remains fixed within a time slot and may change over different time slots due to varying data requests.

Active UE $k$ intends to transmit a $b$-bit binary message $\bm{m}_k$ to the BS. Based on the mechanism of UURA, all UEs share the same codebook. To reduce the implementation complexity of receiver, the $b$-bit message $\bm{m}_k$ of the $k$-th $(k \in \mathcal{K}_a)$ active UE is divided into $L$  short sub-blocks of length $J$ with $b=LJ$, such that the size of common codebook can be small. Meanwhile, a time slot is partitioned into $L$ sub-slots of $n_0$ symbol transmissions each.
Each active UE maps its sub-blocks to codewords based on a given codeword selection scheme.
To be specific, the codebook is given by $\textbf{C}=[\bm{c}_1,...,\bm{c}_{2^J}]\in \mathbb{C}^{n_0\times2^J}$ with each column $\{\bm{c}_i\in \mathbb{C}^{n_0\times1},\ i\in [1:2^J]\}$ representing a codeword. In the $l$-th ($l\in [1,L]$) sub-slot, the $l$-th sub-block of active UE $k$ ($k\in \mathcal{K}_a$) is mapped to integer $i_{k,l}$ and the $i_{k,l}$-th codeword $\bm{c}_{i_{k,l}}$ will be transmitted. For a $J$-bit binary sub-block, there are $2^J$ possible combinations, thus $i_{k,l}\in [1,2^J]$. The mapping will not affect the error results. Herein, the list of active codewords in the $l$-th sub-slot is represented by $\mathcal{L}_l=\{i_{k,l} : k \in \mathcal{K}_a, l\in[1,L]\}$.

After mapping, all active UEs simultaneously send their codewords to the BS over $L$ sub-slots in order. Hence, the received signal $\textbf{Y}_l \in \mathbb{C}^{n_0 \times M}$ at the BS in the $l$-th sub-slot can be expressed as
\begin{eqnarray}\label{model0}
\textbf{Y}_l=\sum\limits_{k\in \mathcal{K}_a}\sqrt{\tilde{g}_k}\bm{c}_{i_{k,l}}\bm{h}^T_{k,l}+\textbf{W}_l
=\textbf{C}\boldsymbol{\Theta}_l\tilde{\textbf{G}}^{1/2}\textbf{H}_l+\textbf{W}_l,
\end{eqnarray}
where $\bm{c}_{i_{k,l}}$ denotes the codeword sent by the $k$-th active UE in the $l$-th sub-slot and satisfies $\|\bm{c}_{i_{k,l}}\|^2=n_0P_k$ with $P_k$ being the per-symbol transmit power. $\bm{h}_{k,l} \in \mathbb{C}^{M \times 1}$ is the small-scale fading vector with i.i.d elements of zero mean and unit variance. It changes independently over sub-slots. Matrix $\textbf{H}_l=[\bm{h}_{1,l},...,\bm{h}_{K_{\rm tot},l}]^T$. $\tilde{\textbf{G}}={\rm diag}(\tilde{g}_1,...,\tilde{g}_{K_{\rm tot}})$ is a diagonal matrix, where the large-scale fading coefficient $\tilde{g}_k$ remains constant throughout the entire time slot. This is the underlying reason for the correlation observed among codewords transmitted by the same active UE. $\textbf{W}_l$ represents the additive white Gaussian noise (AWGN) with zero mean and variance $\sigma^2$.
$\boldsymbol{\Theta}_l\in \mathbb{C}^{2^J\times K_{\rm tot}}$ is the binary indicator matrix. Its $(j,k)$-th element $\theta_l^{j,k}=1$ if $i_{k,l}=j$, otherwise, $\theta_l^{j,k}=0$.
By arrangement, we let codeword state matrix $\textbf{X}_l=\boldsymbol{\Theta}_l\tilde{\textbf{G}}^{1/2}\textbf{H}_l\in\mathbb{C}^{2^J\times M}$, its $j$-th row follows a Bernoulli-Gaussian distribution, i.e., $\forall j\in[1,2^J]$,
\begin{equation}
\bm{x}_{j,l}\sim\left\{
\begin{array}{ll}
\mathcal{CN}(\bm{0},\gamma_{j,l}\textbf{I}),j \in \mathcal{L}_l\\%
\bm{0},\ {\rm otherwise}
\end{array}\right.,
\end{equation}
where $\gamma_{j,l}=\sum_{k\in\mathcal{K}_a}\theta_l^{j,k}\tilde{g}_k$.
Define codeword activity vector $\boldsymbol{\gamma}_l=[\gamma_{1,l},...,\gamma_{2^J,l}]$ and codeword activity matrix $\boldsymbol{\Gamma}_l={\rm diag}(\boldsymbol{\gamma}_l)\in\mathbb{C}^{2^J\times 2^J}$,
then, \eqref{model0} can be rewritten more compactly as
\begin{eqnarray}\label{model}
\textbf{Y}_l=\textbf{C}\boldsymbol{\Gamma}_l^{1/2}\widetilde{\textbf{H}}_l+\textbf{W}_l,
\end{eqnarray}
where $\widetilde{\textbf{H}}_l$ has the i.i.d entries $\sim\mathcal{CN}(0,1)$. In this case, the list of active codewords can be equivalently expressed as $\mathcal{L}_l=\{j : \gamma_{j,l}\neq 0, l\in[1,L]\}$.

\begin{figure}[h] \centering
\includegraphics [width=0.47\textwidth] {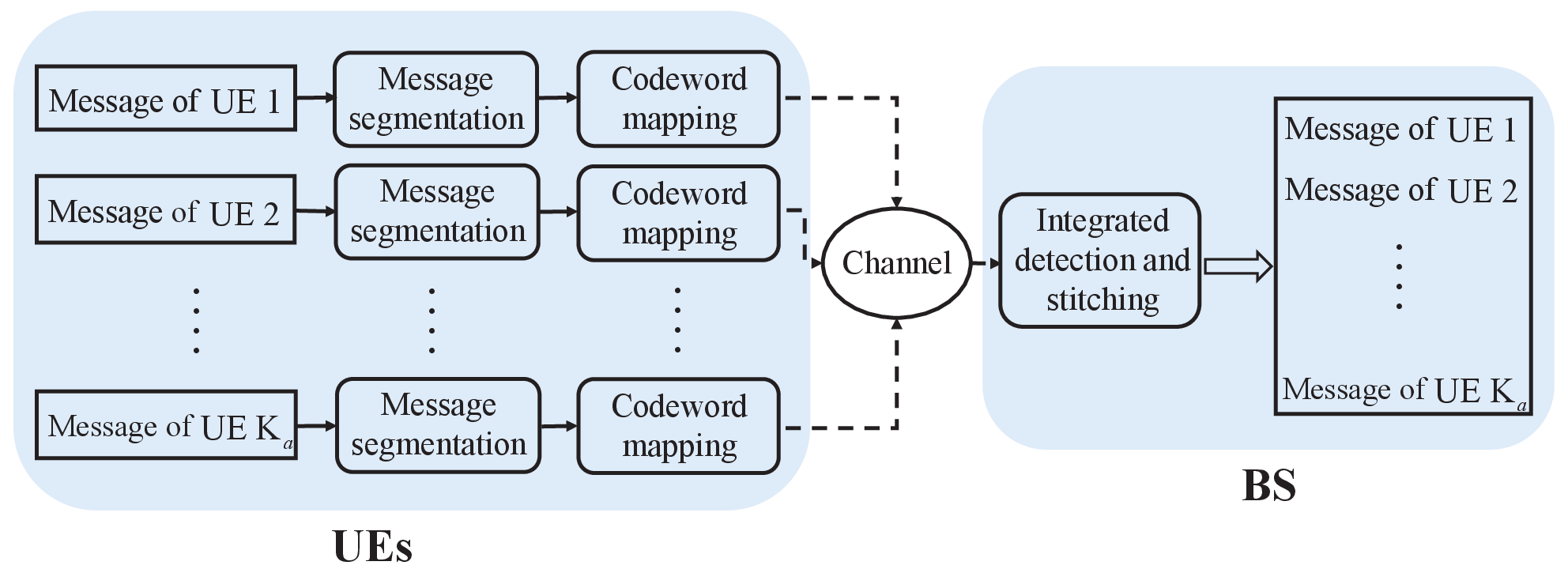}
\caption {The proposed UURA framework.}
\label{Fig0}
\end{figure}

With the received signal of each sub-slot, the BS detects subslot-wise active codewords $\bm{c}_{i_{k,l}}$ of symbol length $n_0$ and stitches the codewords across sub-slots to recover original messages $\bm{m}_k$ sub-block by sub-block for all active UEs. The block diagram of the above UURA framework is shown in Fig. \ref{Fig0}. In the following section, based on the above model setting, we design a decoding algorithm integrating codeword detection and stitching for the BS to achieve this goal in a timely manner.

\section{MIG-aided Integrated decoding of UURA}
In this section, we design an integrated decoding algorithm to recover the original messages according to the noisy received signals at the BS. Specifically, in each sub-slot, the active codewords are detected based on observation $\textbf{Y}_l$ and common codebook $\textbf{C}$ and simultaneously stitched with codewords sent by the same UE in the previous sub-slot by leveraging MIG, such that a portion of the original messages can be obtained timely. The details of the decoding algorithm are provided below.

\subsection{Preliminaries}\label{subsec:MIG}
First, a brief overview of the MIG theory is provided \cite{MIG5}. MIG utilizes geometric methods to tackle signal processing problems involving matrix-valued data on matrix manifolds. By employing geometric modeling, MIG is capable of addressing various tasks, including data clustering and parameter estimation, etc.

Now, let us introduce some fundamental concepts of MIG theory that are relevant to this paper.
Define the Hermitian positive definition (HPD) matrix manifold as
\begin{align}
\mathcal{H}^n_+=\{\textbf{A}\in\mathbb{C}^{n\times n}|\textbf{A}=\textbf{A}^H,\textbf{A}\succ 0\}.
\end{align}
The manifold $\mathcal{H}^n_+$ represents a collection of matrices satisfying the HPD property, with such matrices being regarded as data points on the manifold. Performing matrix operations and geometric analysis within this collection preserves the structural properties of matrices. 
Within the manifold $\mathcal{H}^n_+$, two matrix-valued data points can be connected by multiple curves that describe the paths of variation between them. These curves play a crucial role in geometric analysis and optimization. The shortest path among these curves is known as the geodesic distance. Geodesic distances can be measured using different metrics, capturing the geometric disparities between matrix-valued data points.

There exist various distance measurements, and choosing an appropriate one is crucial when designing an efficient codeword stitching method.
In this paper, the geodesic distance on HPD matrix manifold $\mathcal{H}^n_+$, denoted by $\mathcal{D}(\cdot)$, is defined as the logarithmic Euclidean distance
\begin{align}
\mathcal{D}^2(\textbf{A},\textbf{B})=\|{\rm Log}(\textbf{A}) - {\rm Log}(\textbf{B})\|_F^2.
\end{align}
${\rm Log}(\cdot)$ is the logarithm of HPD matrix, which can be expressed as the following power series
\begin{align}
{\rm Log}(\textbf{A})=-\sum_{k=1}^{\infty}\frac{(\textbf{I}-\textbf{A})^k}{k},\|\textbf{I}-\textbf{A}\|<1.
\end{align}
Similar to the algebraic mean, the geometric center of a set of matrix-valued data $\{\textbf{A}_1,\textbf{A}_2,...,\textbf{A}_N\}$ on the manifold $\mathcal{H}^n_+$ is given by the solution to the following optimization problem
\begin{align}
\overline{\textbf{A}}={\rm arg}\min_{\textbf{A}\in\mathcal{H}^n_+}\sum_{i=1}^N\mathcal{D}^2(\textbf{A},\textbf{A}_i).
\end{align}

For the considered UURA, it is evident that the covariance matrices of the non-zero rows of the codeword state matrix $\textbf{X}_l$, namely codeword covariance matrices, belong to the manifold $\mathcal{H}^M_+$, i.e.,
\begin{align}
\textbf{R}_j^l=\gamma_{j,l}\textbf{I}\in\mathcal{H}^M_+,j\in\mathcal{L}_l.
\end{align}
Thereby, the geodesic distance can be utilized to determine the similarities or dissimilarities between codeword covariance matrices for the purpose of codeword concatenation.
Based on this, the fundamental idea behind employing MIG for codeword concatenation is as follows: within a set of matrix-valued data $\mathcal{R}=\{\textbf{R}_j^l, j\in\mathcal{L}_l, l\in[1,L]\}$, the codeword covariance matrices in the subset $\mathcal{R}_k^L=\{\textbf{R}_{i_{k,1}}^1, \textbf{R}_{i_{k,2}}^2,...,\textbf{R}_{i_{k,L}}^L\}$, which are determined by $L$ codewords transmitted by the same active UE $k$, should exhibit a smaller geodesic distance between them. This is because they contain the channel fading information of the same active UE.
Consequently, the process of codeword stitching can be realized by finding and continuously updating the geometric center of the subset $\mathcal{R}_k^{l-1},k\in\mathcal{K}_a$, which is defined as
\begin{align}\label{geocenter}
\textbf{G}_{k}^{l}=\left\{
\begin{array}{ll}
\exp\left[\frac{1}{l}\sum_{\textbf{R}_j^i\in\mathcal{R}_k^{l}}{\rm Log}(\textbf{R}_j^i)\right],&l=2,...,L\\
\textbf{R}_j^1,&l=1
\end{array}\right..
\end{align}
Additionally, the aim is to minimize the geodesic distance between the codeword covariance matrix $\textbf{R}_{i_{k,l}}^l$ and the geometric center $\textbf{G}_{k}^{l-1}$ of the corresponding subset to which it belongs. Note that \eqref{geocenter} represents the theoretical definition of the geometric center and its calculation does not require to keep all true $\mathbf{R}_j^l$ in the subsequent algorithm recursions.

In what follows, a detailed explanation of the problem formulation and the solution for the decoding algorithm will be presented according to the above principles.

\subsection{Problem Formulation}
As mentioned earlier, the key of the decoding design is the integrated codeword detection and stitching. Put simply, the objective is to recover the unknown codeword activity vector $\boldsymbol{\gamma}_l$ in the system model \eqref{model}. Additionally, the active codewords from the estimated set of the $l$-th sub-slot, determined by the estimated vector $\hat{\boldsymbol{\gamma}}_l$, need to be concatenated with the codewords from the previous sub-slots.
Essentially, we require to identify a total of $K_a\times L$ active codewords and divide them into $K_a$ classes based on whether they are sent from the same active UE. By stitching together the active codewords within each class in chronological order, we can reconstruct the original messages from all active UEs.

\begin{figure*}[!ht] \centering
\includegraphics [width=0.77\textwidth] {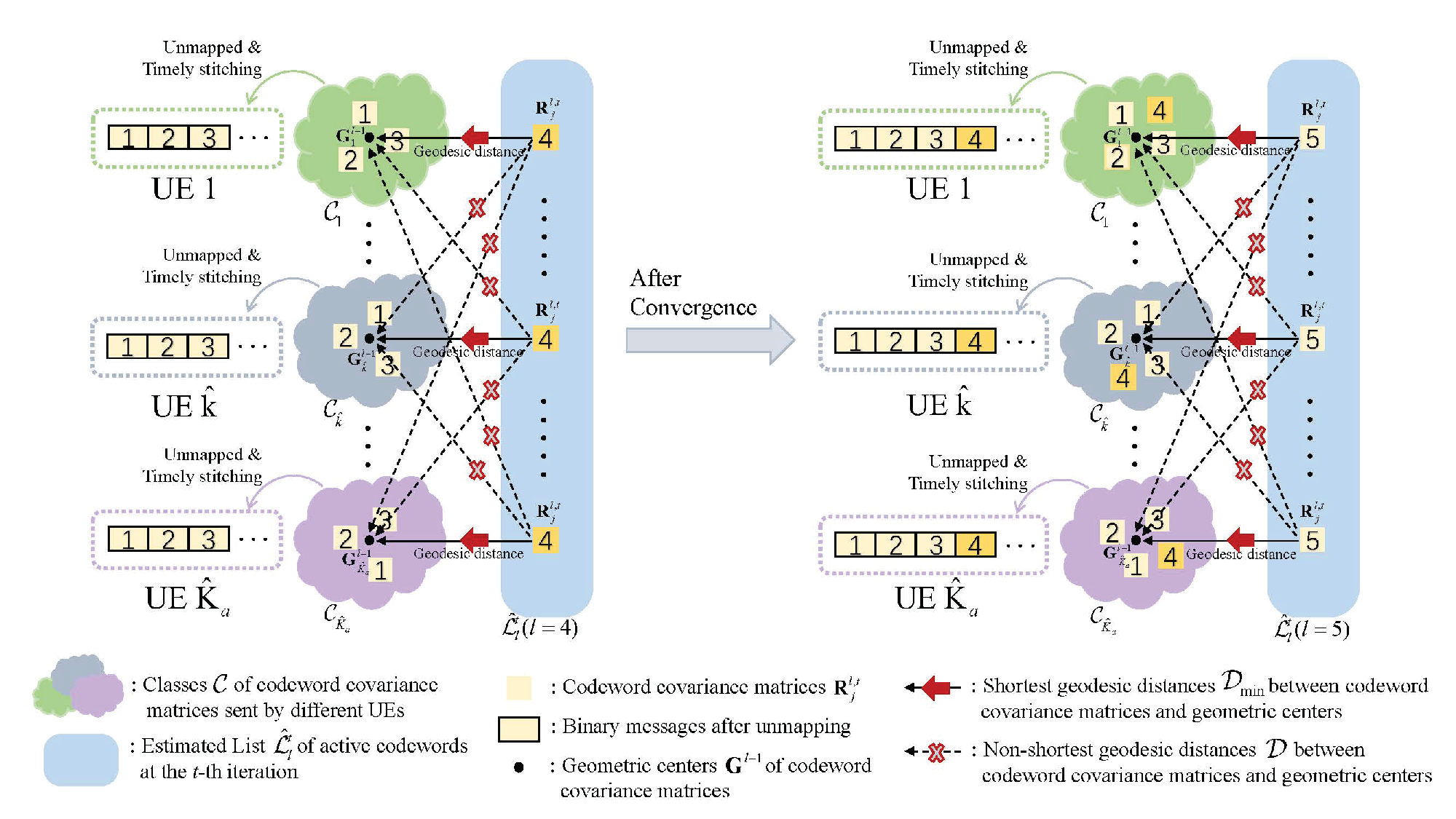}
\caption {The illustration of the proposed integrated decoding algorithm. The codeword detection and stitching are performed simultaneously during the solving of P1 by the receiver. Specifically, in the $l$-th sub-slot (e.g., $l=4$ or $5$ shown in the figure), the temporary estimated list $\hat{\mathcal{L}}_l^t$ of active codewords is obtained at the beginning of each iteration. For each active codeword $j\in\hat{\mathcal{L}}_l^t$, the class $\mathcal{C}_{\hat{k}}$ with the shortest geodesic distance is determined. Then, the iteration value of the variable $\boldsymbol{\gamma}_l^{t}$ is updated by minimizing the geodesic distance between the covariance matrix $\textbf{R}_j^{l,t}$ of codeword $j$ and the geometric center $\textbf{G}_{\hat{k}}^{l-1}$ of the class $\mathcal{C}_{\hat{k}}$. After convergence, the active codewords in the $l$-th sub-slot are detected and grouped into $\hat{K}_a$ classes. By unmapping the codeword indices in each class to binary bits, the sub-messages can be timely stitched in chronological order.}
\label{Fig1}
\end{figure*}

Taking into account the characteristics of the UURA framework, we formulate the problem of integrated codeword detection and stitching, denoted as  $\min_{\boldsymbol{\gamma}_l\in\mathbb{R}_+}\xi(\boldsymbol{\gamma}_l)$, by associating an error minimization estimator $\min_{\boldsymbol{\gamma}_l\in\mathbb{R}_+}f(\boldsymbol{\gamma}_l)$ with regularized sparsity-promoting term $g(\boldsymbol{\gamma}_l)$ and stitching-promoting term $\phi(\boldsymbol{\gamma}_l)$.

Specifically, we first model the error minimization estimator using a well-known covariance-based approach that leverages the maximum likelihood (ML) principle. Herein, the codeword activity vector can be detected from the noisy observations $\textbf{Y}_l$ and the common codebook $\textbf{C}$. To be more precise, considering the system model in \eqref{model}, each column of $\textbf{Y}_l$, denoted by $\bm{y}_{m,l},m\in[1,M]$, can be regarded as an independent sample following complex Gaussian distribution, i.e.,
\begin{align}
\bm{y}_{m,l}\sim\mathcal{CN}(\bm{0},\boldsymbol{\Psi}_l),
\end{align}
where covariance matrix $\boldsymbol{\Psi}_l={\rm E}\{\bm{y}_{m,l}\bm{y}_{m,l}^H\}=\textbf{C}\boldsymbol{\Gamma}_l\textbf{C}^H+\sigma^2\textbf{I}$.
As a result, the likelihood function of $\textbf{Y}_l$ given $\boldsymbol{\gamma}_l$ can be expressed as
\begin{align}
P(\textbf{Y}_l|\boldsymbol{\gamma}_l)=\frac{1}{|\boldsymbol{\Psi}_l|^M}{\rm exp}\left[-{\rm tr}(\boldsymbol{\Psi}_l^{-1}\hat{\boldsymbol{\Psi}}_l^Y)\right],
\end{align}
where sample covariance matrix of the received signals $\hat{\boldsymbol{\Psi}}_l^Y=\frac{1}{M}\textbf{Y}_l\textbf{Y}_l^H$.
Based on this, the error minimization estimator can be presented as a problem of minimizing the negative log-likelihood function, which is given by
\begin{align}\label{fx}
\min_{\boldsymbol{\gamma}_l\in\mathbb{R}_+} f(\boldsymbol{\gamma}_l)&=-\ln(P(\textbf{Y}_l|\boldsymbol{\gamma}_l))\nonumber\\
&=\ln|\boldsymbol{\Psi}_l|+{\rm tr}(\boldsymbol{\Psi}_l^{-1}\hat{\boldsymbol{\Psi}}_l^Y).
\end{align}

Notice that in the first sub-slot, only codeword detection needs to be performed, and codeword stitching is not required.
Therefore, we can directly address the decoding problem in the first sub-slot using the above ML-based detection approach, without incorporating any regularization terms. That is,
\begin{align}\label{P0}
{\rm P0:}\min_{\boldsymbol{\gamma}_l\in\mathbb{R}_+} \xi(\boldsymbol{\gamma}_l)=f(\boldsymbol{\gamma}_l), l=1.
\end{align}
Solving P0 allows us to obtain the estimated codeword activity vector $\hat{\boldsymbol{\gamma}_1}$.
In this case, the estimated set of active codewords can be judged in an element-wise way, i.e., $\hat{\mathcal{L}}_1=\{j:\hat{\gamma}_{j,1}>\varpi_1,j\in[1,2^J]\}$ with $\varpi_1>0$ being a suitable predefined threshold and $\hat{\gamma}_{j,1}$ being the $j$-th element of $\hat{\boldsymbol{\gamma}}_1$.
It is worth mentioning that the optimal threshold, which achieves the minimum error probability, can be set either through exhaustive search or with the assistance of channel strength and the asymptotic Gaussian characteristics of ML estimation error, provided that the true fading coefficients are available \cite{SP}.
At this time, the estimated number of active UEs can be derived by $\hat{K}_a=|\hat{\mathcal{L}}_1|$.
Based on this, we divide $\hat{K}_a$ codewords in $\hat{\mathcal{L}}_1$ into $\hat{K}_a$ different classes, denoted by $\{\mathcal{C}_{\hat{k}},\hat{k}\in[1,\hat{K}_a]\}$, and define the initial geometric center of codeword covariance matrices within each class as $\textbf{G}_{\hat{k}}^1=\hat{\textbf{R}}_j^1=\hat{\gamma}_{j,1}\textbf{I}$, $j\in\hat{\mathcal{L}}_1\cap\mathcal{C}_{\hat{k}}$.

\emph{Remark:} In scenarios where detection errors are present, it is possible for $\hat{K}_a$ to deviate from the true $K_a$. Consequently, the number of classes utilized for codeword concatenation may increase or decrease, leading to missed detection of messages from active UEs or the erroneous detection of messages that were not transmitted.
Fortunately, the employment of the ML-based method to solve P0 can achieve high detection accuracy under favorable power conditions and reduce the decoding errors at the source \cite{UN9}. Hence, it is temporarily assumed that $\hat{K}_a$ is accurate.
As a supplement, we conduct experiments to evaluate the impact of the accuracy of $\hat{K}_a$ on the overall performance of the decoding algorithm and justify this assumption in Section \ref{sec:sim}.

Then, let us proceed with the design of regularization terms for the integrated decoding in subsequent sub-slots. Due to the sporadic activity of IoT UEs, the codeword activity vector $\boldsymbol{\gamma}_l$ is sparse. Promoting sparsity in the codeword activity vector not only aids in detecting $\boldsymbol{\gamma}_l$ but also reduces the density of codeword covariance matrices on the matrix manifold. This increases the dissimilarity between matrices, facilitating the codeword concatenation. In general, minimizing $l_0$-norm of a vector can be employed to encourage its element sparsity. However, $l_0$-norm is non-convex and non-differentiable because it directly counts the number of non-zero elements. Therefore, the convex relaxation of $l_0$-norm, i.e., $l_1$-norm, is utilized and the sparsity-promoting term is expressed as
\begin{align}\label{gx}
\min_{\boldsymbol{\gamma}_l\in\mathbb{R}_+} g(\boldsymbol{\gamma}_l)=\|\boldsymbol{\gamma}_l\|_1.
\end{align}
Apart from promoting sparsity, an essential aspect of integrated decoding is stitching promotion. To achieve it, we extract codeword covariance matrices as the clustering features and measure the geometric similarities between these matrices using MIG as the stitching basis. Following the ideas introduced in subsection \ref{subsec:MIG}, we design a stitching-promoting term to minimize the geodesic distance between the covariance matrices $\textbf{R}_j^l=\gamma_{j,l}\textbf{I}$ of codewords sent by the same active UE, which is written as
\begin{align}\label{phix}
\min_{\boldsymbol{\gamma}_l\in\mathbb{R}_+} \phi(\boldsymbol{\gamma}_l)=\sum_{j\in\mathcal{L}_l}\min_{\hat{k}}\|{\rm Log}(\textbf{G}_{\hat{k}}^{l-1})-{\rm Log}(\textbf{R}_j^l)\|_F^2
\end{align}
with $\hat{k}\in[1,\hat{K}_a]$. Note that if we were to adopt a temporally or spatially correlated channel model, the design of the stitching-promoting term could benefit from additional available information. However, it is necessary to consider the challenges associated with algorithmic processing in such cases.
By incorporating these regularization terms, the efficient integrated decoding can be realized by promoting sparsity in the codeword activity vector and encouraging the proximity of the codeword covariance matrices with geometric similarities. These terms play a critical role in enhancing the overall decoding process. To this end, we combine the local error minimization estimator \eqref{fx} with two regularization terms \eqref{gx} and \eqref{phix}, resulting in the following optimization problem
\begin{align}\label{P1}
{\rm P1:}\min_{\boldsymbol{\gamma}_l\in\mathbb{R}_+} \xi(\boldsymbol{\gamma}_l)=f(\boldsymbol{\gamma}_l)+\alpha g(\boldsymbol{\gamma}_l)+\beta\phi(\boldsymbol{\gamma}_l), l\in[2,L],
\end{align}
where $\alpha>0,\beta>0$ are the penalty parameters.
Next, we will develop a decoding algorithm to solve P1 and obtain the estimation $\hat{\boldsymbol{\gamma}_l}$, which jointly minimizes the costs involved in \eqref{P1}, for each sub-slot.

\subsection{Algorithm Recursions}
As we know, the traditional gradient descent method is suitable for solving minimization problems involving differentiable convex functions. In this paper, the first term of P1, $f(\boldsymbol{\gamma}_l)$, is geodesically convex and differentiable \cite{UN9}. However, the second term $g(\boldsymbol{\gamma}_l)$ is non-differentiable and the third term $\phi(\boldsymbol{\gamma}_l)$ is the sum of non-smooth functions, even though they are convex. Fortunately, proximal gradient method can efficiently address the non-smooth convex problems \cite{Proxi,Proxi1,Proxi2}.
Moreover, the variable $\boldsymbol{\gamma}_l$ for different sub-slots are coupled with each other. In this context, we aim to design an algorithm that independently detects codeword activity vectors for each sub-slot. By leveraging the detection results from the preceding sub-slots, we can improve the codeword detection performance in subsequent sub-slots and simultaneously accomplish codeword stitching. In what follows, we introduce the implementation process of the proposed decoding algorithm based on the proximal gradient method. Specifically, the proximal gradient descent for solving P1 can be expressed as the following iteration
\begin{align}\label{iter0}
\boldsymbol{\gamma}_l^{t+1}&=\boldsymbol{\gamma}_l^t-\eta_l^t\nabla f(\boldsymbol{\gamma}_l^t)-\alpha\eta_l^t\partial g(\boldsymbol{\gamma}_l^{t+1})-\beta\eta_l^t\partial\phi(\boldsymbol{\gamma}_l^{t+1})\nonumber\\
&={\rm prox}_{\eta_l^t(\alpha g + \beta\phi)}\left(\boldsymbol{\gamma}_l^t-\eta_l^t\nabla f(\boldsymbol{\gamma}_l^t)\right)
\end{align}
with $\boldsymbol{\gamma}_l^t$ and $\boldsymbol{\gamma}_l^{t+1}$ being the values of $\boldsymbol{\gamma}_l$ at the $t$-th and the $(t+1)$-th iterations. $\eta_l^t$ is the step size of the $t$-th iteration and proximal operator is given by
\begin{align}
x^{t+1}={\rm prox}_{\eta g}(x^t)={\rm arg}\min_x g(x) + \frac{1}{2\eta}\|x-x^t\|_2^2.
\end{align}
$\partial g(\boldsymbol{\gamma}_l^{t+1})$ and $\partial\phi(\boldsymbol{\gamma}_l^{t+1})$ denote the subgradients of functions $g(\boldsymbol{\gamma}_l)$ and $\phi(\boldsymbol{\gamma}_l)$ at $\boldsymbol{\gamma}_l^{t+1}$. They are converted into the proximal operation ${\rm prox}_{\eta_l^t(\alpha g + \beta\phi)}$ so that the iteration in \eqref{iter0} can be rewritten as an expression of $\boldsymbol{\gamma}_l^{t+1}$ solely in terms of $\boldsymbol{\gamma}_l^t$.
$\nabla f(\boldsymbol{\gamma}_l^t)$ is the gradient of function $f(\boldsymbol{\gamma}_l)$ at $\boldsymbol{\gamma}_l^t$, it can be calculated according to the well known Sherman-Morrison rank-1 update property \cite{SMupdate}. To be specific,
let $\boldsymbol{\Psi}_l^t=\boldsymbol{\Psi}_{j,l}^t+\gamma_{j,l}\bm{c}_j\bm{c}_j^H$ with $\bm{c}_j$ being the $j$-th column of codebook matrix $\textbf{C}$, we have
\begin{align}
(\boldsymbol{\Psi}_{j,l}^t + \gamma_{j,l}^t\bm{c}_j\bm{c}_j^H)^{-1}=(\boldsymbol{\Psi}_{j,l}^t)^{-1} - \frac{\gamma_{j,l}^t(\boldsymbol{\Psi}_{j,l}^t)^{-1}\bm{c}_j\bm{c}_j^H(\boldsymbol{\Psi}_{j,l}^t)^{-1}}{1 + \gamma_{j,l}^t\bm{c}_j^H(\boldsymbol{\Psi}_{j,l}^t)^{-1}\bm{c}_j}
\end{align}
and
\begin{align}
|\boldsymbol{\Psi}_{j,l}^t + \gamma_{j,l}^t\bm{c}_j\bm{c}_j^H|=\left(1 + \gamma_{j,l}^t\bm{c}_j^H(\boldsymbol{\Psi}_{j,l}^t)^{-1}\bm{c}_j\right)|\boldsymbol{\Psi}_{j,l}^t|.
\end{align}
Thus, $f(\boldsymbol{\gamma}_l^t)$ can be rewritten as
\begin{align}
&f(\boldsymbol{\gamma}_l^t)=\ln\left(1+\gamma_{j,l}^t\bm{c}_j^H(\boldsymbol{\Psi}_{j,l}^t)^{-1}\bm{c}_j\right)+\ln|\boldsymbol{\Psi}_{j,l}^t|\nonumber\\
&+{\rm tr}\left((\boldsymbol{\Psi}_{j,l}^t)^{-1}\hat{\boldsymbol{\Psi}}_l^Y\right)-\frac{\gamma_{j,l}^t\bm{c}_j^H(\boldsymbol{\Psi}_{j,l}^t)^{-1} \hat{\boldsymbol{\Psi}}_l^Y(\boldsymbol{\Psi}_{j,l}^t)^{-1}\bm{c}_j}{1+\gamma_{j,l}^t\bm{c}_j^H(\boldsymbol{\Psi}_{j,l}^t)^{-1}\bm{c}_j}.
\end{align}
Taking the derivative of the function $f(\boldsymbol{\gamma}_l^t)$ with respect to its element $\gamma_{j,l}^t$, we have
\begin{align}
\frac{\partial f(\boldsymbol{\gamma}_l^t)}{\partial\gamma_{j,l}^t}&=\frac{\bm{c}_j^H(\boldsymbol{\Psi}_{j,l}^t)^{-1}\bm{c}_j}{1+\gamma_{j,l}^t\bm{c}_j^H(\boldsymbol{\Psi}_{j,l}^t)^{-1}\bm{c}_j}\nonumber\\ &\quad\quad\quad - \frac{\bm{c}_j^H(\boldsymbol{\Psi}_{j,l}^t)^{-1} \hat{\boldsymbol{\Psi}}_l^Y(\boldsymbol{\Psi}_{j,l}^t)^{-1}\bm{c}_j}{(1+\gamma_{j,l}^t\bm{c}_j^H(\boldsymbol{\Psi}_{j,l}^t)^{-1}\bm{c}_j)^2}.
\end{align}
Consequently, the gradient $\nabla f(\boldsymbol{\gamma}_l^t)$ can be presented as
\begin{align}\label{graf}
\nabla f(\boldsymbol{\gamma}_l^t)=\left[\frac{\partial f(\boldsymbol{\gamma}_l^t)}{\partial\gamma_{1,l}^t},...,\frac{\partial f(\boldsymbol{\gamma}_l^t)}{\partial\gamma_{2^J,l}^t}\right]^T.
\end{align}

Next, let's move on to solve the proximal operation ${\rm prox}_{\eta_l^t(\alpha g + \beta\phi)}$ in \eqref{iter0}. Yet, it is intractable to directly address the proximal operator of the sum of $\alpha g + \beta\phi$. Fortunately, Douglas-Rachford splitting technique can be exploited to alternately update and separately handle two proximal operators \cite{Proxi}. Specifically, we define the error signal $\bm{u}^t=\partial\phi(\boldsymbol{\gamma}_l^{t+1})$ and introduce an auxiliary variable $\bm{z}^t$. By doing so, the gradient descent operation of $\phi(\cdot)$ can be incorporated into the input of the proximal operation of $g(\cdot)$. This allows us to handle the two regularization terms separately and update them iteratively. That is,
\begin{align}\label{iter1}
\bm{z}^t={\rm prox}_{\eta_l^t(\alpha g)}\left(\boldsymbol{\gamma}_l^t-\eta_l^t\nabla f(\boldsymbol{\gamma}_l^t)-\eta_l^t\beta\bm{u}^t\right).
\end{align}
According to the definition of proximal operator, $\bm{z}^t$ can be derived as
\begin{align}
\bm{z}^t&={\rm arg}\min_{\boldsymbol{\gamma}_l}\left(g(\boldsymbol{\gamma}_l)+\frac{1}{2\alpha\eta_l^t}\|\boldsymbol{\gamma}_l-\bm{a}^t\|_2^2\right)\nonumber\\
&={\rm arg}\min_{\boldsymbol{\gamma}_l}\left(\alpha\eta_l^tg(\boldsymbol{\gamma}_l)+\|\boldsymbol{\gamma}_l-\bm{a}^t\|_2^2\right),
\end{align}
where $\bm{a}^t=\boldsymbol{\gamma}_l^t-\eta_l^t\nabla f(\boldsymbol{\gamma}_l^t)-\eta_l^t\beta\bm{u}^t$. Taking the derivative of the above function in bracket and making it equal to $0$, we have
\begin{align}\label{der1}
\alpha\eta_l^t\frac{\partial g(\boldsymbol{\gamma}_l)}{\partial\boldsymbol{\gamma}_l} + \boldsymbol{\gamma}_l-\bm{a}^t=0.
\end{align}
Substituting $g(\boldsymbol{\gamma}_l)=\|\boldsymbol{\gamma}_l^t\|_1$ into \eqref{der1}, we obtain
\begin{align}
\bm{z}^t=\bm{a}^t-\alpha\eta_l^t\left[{\rm sign}(\gamma_{1,l}^t),...,{\rm sign}(\gamma_{2^J,l}^t)\right]^T
\end{align}
or its equivalent element-wise form, soft threshold operator,
\begin{align}\label{zt}
z_i^t=\left\{
\begin{array}{lcl}
a_i^t-\alpha\eta_l^t,&a_i^t&\!\!>\alpha\eta_l^t\\
0,&|a_i^t|&\!\!\leq\alpha\eta_l^t\\
a_i^t+\alpha\eta_l^t,&a_i^t&\!\!<-\alpha\eta_l^t
\end{array}\right..
\end{align}
Similarly, the auxiliary variable $\bm{z}^t$ can also be involved in the input of proximal operation of $\phi(\cdot)$ based on the Douglas-Rachford splitting rule. That is,
\begin{align}\label{iter2}
\boldsymbol{\gamma}_l^{t+1}={\rm prox}_{\eta_l^t(\beta\phi)}\left(\bm{z}^t+\eta_l^t\beta\bm{u}^t\right).
\end{align}
According to the definition of proximal operator, $\boldsymbol{\gamma}_l^{t+1}$ can be derived as
\begin{align}
\boldsymbol{\gamma}_l^{t+1}&={\rm arg}\min_{\boldsymbol{\gamma}_l}\left(\phi(\boldsymbol{\gamma}_l)+\frac{1}{2\beta\eta_l^t}\|\boldsymbol{\gamma}_l-\bm{b}^t\|_2^2\right)\nonumber\\
&={\rm arg}\min_{\boldsymbol{\gamma}_l}\left(\beta\eta_l^t\phi(\boldsymbol{\gamma}_l)+\|\boldsymbol{\gamma}_l-\bm{b}^t\|_2^2\right),
\end{align}
where $\bm{b}^t=\bm{z}^t+\eta_l^t\beta\bm{u}^t$. Taking the derivative of the above function in bracket and making it equal to $0$, we have
\begin{align}\label{der2}
\beta\eta_l^t\frac{\partial \phi(\boldsymbol{\gamma}_l)}{\partial\boldsymbol{\gamma}_l} + \boldsymbol{\gamma}_l-\bm{b}^t=0.
\end{align}
Substituting $\phi(\boldsymbol{\gamma}_l)=\sum_{j\in\hat{\mathcal{L}}_l^t}\min_{\hat{k}}\|{\rm Log}(\textbf{G}_{\hat{k}}^{l-1})-{\rm Log}(\textbf{R}_j^{l,t})\|_F^2$ into \eqref{der2}, we obtain
\begin{align}\label{gamatp1}
\boldsymbol{\gamma}_l^{t+1}=\bm{b}^t-\beta\eta_l^t\left[\frac{\partial\phi(\boldsymbol{\gamma}_l^t)}{\partial\gamma_{1,l}^t},...,\frac{\partial\phi(\boldsymbol{\gamma}_l^t)}{\partial\gamma_{2^J,l}^t}\right]^T,
\end{align}
where $\frac{\partial\phi(\boldsymbol{\gamma}_l^t)}{\partial\gamma_{j,l}^t}=\left(\ln(\gamma_{j,l}^t)-r_{k_{j*}^t,l-1}\right)/\gamma_{j,l}^t$ and $r_{k_{j*}^t,l-1}=\frac{1}{l-1}\sum_{i=1}^{l-1}\ln(\hat{\gamma}_{j,l}),j\in\hat{\mathcal{L}}_i\cap\mathcal{C}_{k_{j*}^t}$.
$\hat{\mathcal{L}}_l^t$ is the temporary estimated list of active codewords in the $t$-th iteration and
$k_{j*}^t={\rm arg}\min_{\hat{k}}\|{\rm Log}(\textbf{G}_{\hat{k}}^{l-1})-{\rm Log}(\textbf{R}_j^{l,t})\|_F^2$ is the temporary class index with $\textbf{R}_j^{l,t}=\gamma_{j,l}^t\textbf{I}$.

After getting $\boldsymbol{\gamma}_l^{t+1}$, the covariance matrix $\boldsymbol{\Psi}_l$ and error signal $\bm{u}^t$ need to be updated as
\begin{align}
\boldsymbol{\Psi}_l^{t+1}=\boldsymbol{\Psi}_l^t + (\gamma_{j,l}^{t+1}-\gamma_{j,l}^t)\bm{c}_j\bm{c}_j^H
\end{align}
and
\begin{align}
\bm{u}^{t+1}=\bm{u}^t+\frac{1}{\eta_l^t\beta}(\bm{z}^t-\boldsymbol{\gamma}_l^{t+1}),
\end{align}
which will be used in the next iteration.
Note that codeword activity vector $\boldsymbol{\gamma}_l^t$ and auxiliary variable $\bm{z}^t$ are alternately updated during the iterations and will converge to the same value. Meanwhile, the error signal $\bm{u}^t$ will gradually approach zero \cite{Proxi}.

\begin{algorithm}
\caption{: Integrated Decoding of UURA}
\label{alg1}
{\bf Input:}
The codebook matrix $\textbf{C}$, the received signals $\{\textbf{Y}_l\}_{l=1}^L$, sample covariance matrices $\{\hat{\boldsymbol{\Psi}}_l^Y=\frac{1}{M}\textbf{Y}_l\textbf{Y}_l^H\}_{l=1}^L$, step size $\{\eta_l^0\}_{l=1}^L$, decision threshold $\varpi_1$, convergence accuracy $\varpi_2$ and the maximum number of iteration $T_{\rm iter}$.\\
{\bf Output:} The estimated message.
\begin{algorithmic}[1]
\FOR{$l=1:L$}
\STATE{Initialize $\hat{\boldsymbol{\gamma}}_l^0=\textbf{0}$;}
\STATE{Compute the initial estimation $\boldsymbol{\gamma}_l^{ML}$ by solving P0;}
\IF{$l=1$}
  \STATE{Let $\hat{\boldsymbol{\gamma}}_l=\boldsymbol{\gamma}_l^{ML}$;}
  \STATE{Judge $\hat{\mathcal{L}}_l=\{j:\hat{\gamma}_{j,l}>\varpi_1,j\in[1,2^J]\}$ and compute $\hat{K}_a=|\hat{\mathcal{L}}_l|$;}
  \STATE{Let $\hat{K}_a$ codewords in $\hat{\mathcal{L}}_l$ belong to $\hat{K}_a$ different classes $\mathcal{C}_{\hat{k}}$, $\hat{k}\in[1,\hat{K}_a]$;}
  \STATE{Set initial geometric center of each class $\textbf{G}_{\hat{k}}^l=\hat{\textbf{R}}_j^l=\hat{\gamma}_{j,l}\textbf{I}$, $j\in\hat{\mathcal{L}}_l\cap \mathcal{C}_{\hat{k}}$;}
  \STATE{Output $\hat{\bm{m}}_{\hat{k}}((l-1)J+1:lJ)=\textmd{unmapped}(j)\in \mathbb{B}^{J\times 1},j\in\mathcal{C}_{\hat{k}}\cap\hat{\mathcal{L}}_l,\hat{k}\in[1,\hat{K}_a]$.}
\ELSE
  \STATE{Initialize $\{\boldsymbol{\gamma}_l^t\}_{t=1}=\boldsymbol{\gamma}_l^{ML}$, $\{\boldsymbol{\Psi}_l^t\}_{t=1}=\sigma^2\textbf{I}$ and $\{\bm{u}^t\}_{t=1}=\bm{0}$;}
  \FOR{$t=1:T_{\rm iter}$}
    \STATE{Judge the indices of the $\hat{K}_a$ largest elements of $\boldsymbol{\gamma}_l^t$ into $\hat{\mathcal{L}}_l^t$;}
    \STATE{Compute minimum geodesic distance for $j\in\hat{\mathcal{L}}_l^t$;}
    \STATE{Compute $\nabla f(\boldsymbol{\gamma}_l^t)$ according to \eqref{graf};}
    \STATE{Compute auxiliary variable $\bm{z}^t$ based on \eqref{zt};}
    \STATE{Compute $\boldsymbol{\gamma}_l^{t+1}$ based on \eqref{gamatp1};}
    \STATE{Update $\boldsymbol{\Psi}_l^{t+1}=\boldsymbol{\Psi}_l^t + (\gamma_{j,l}^{t+1}-\gamma_{j,l}^t)\bm{c}_j\bm{c}_j^H$;}
    \STATE{Update $\bm{u}^{t+1}=\bm{u}^t+\frac{1}{\eta_l^t\beta}(\bm{z}^t-\boldsymbol{\gamma}_l^{t+1})$;}
    \IF{$\|\boldsymbol{\gamma}_l^{t+1}-\boldsymbol{\gamma}_l^t\|_2^2/{2^J}<\varpi_2$ or $t=T_{\rm iter}$}
        \STATE{Let $\hat{\boldsymbol{\gamma}}_l=\boldsymbol{\gamma}_l^{t+1}$;}
        \STATE{\textbf{Break;}}
    \ENDIF
  \ENDFOR
  \STATE{Judge the indices of the $\hat{K}_a$ largest elements of $\hat{\boldsymbol{\gamma}}_l$ into $\hat{\mathcal{L}}_l$;}
  \STATE{Compute $k_{j*}^{\rm fin}={\rm arg}\min_{\hat{k}}\|{\rm Log}(\textbf{G}_{\hat{k}}^{l-1})- {\rm Log}(\hat{\textbf{R}}_{j}^{l})\|_F^2$. Let each codeword $j\in\hat{\mathcal{L}}_l$ belongs to the $k_{j*}^{\rm fin}$-th class $\mathcal{C}_{k_{j*}^{\rm fin}}$;}
  \STATE{Update the geometric center of each class $\textbf{G}_{\hat{k}}^l=\exp[\frac{1}{l}\sum_{i=1}^l{\rm Log}(\hat{\textbf{R}}_j^i)]$ with $\hat{\textbf{R}}_j^i=\hat{\gamma}_{j,i}\textbf{I}$, $j\in\hat{\mathcal{L}}_i\cap \mathcal{C}_{\hat{k}}$;}
  \STATE{Output $\hat{\bm{m}}_{\hat{k}}((l-1)J+1:lJ)=\textmd{unmapped}(j)\in \mathbb{B}^{J\times 1},j\in\mathcal{C}_{\hat{k}}\cap\hat{\mathcal{L}}_l,\hat{k}\in[1,\hat{K}_a]$.}
\ENDIF
\ENDFOR
\end{algorithmic}
\end{algorithm}

Once the estimated codeword activity vector $\hat{\boldsymbol{\gamma}}_l$ of the $l$-th sub-slot is acquired after convergence of iterations in \eqref{iter1} and \eqref{iter2}, we select the indices corresponding to the $\hat{K}_a$ largest entries of $\hat{\boldsymbol{\gamma}}_l$ to obtain the estimated list $\hat{\mathcal{L}}_l$ of active codewords in the current sub-slot.
Then, compute the final class index of each active codeword in $\hat{\mathcal{L}}_l$ as $k_{j*}^{\rm fin}={\rm arg}\min_{\hat{k}}\|{\rm Log}(\textbf{G}_{\hat{k}}^{l-1})-{\rm Log}(\hat{\textbf{R}}_j^l)\|_F^2,\hat{k}\in[1,\hat{K}_a], \hat{\textbf{R}}_j^l=\hat{\gamma}_{j,l}\textbf{I}$. This indicates that codeword $j\in\hat{\mathcal{L}}_l$ belongs to class $\mathcal{C}_{k_{j*}^{\rm fin}}$. In other words, the codeword $j\in\hat{\mathcal{L}}_l\cap\mathcal{C}_{k_{j*}^{\rm fin}}$ will be stitched together with the previous codeword $j\in\hat{\mathcal{L}}_{l-1}\cap\mathcal{C}_{k_{j*}^{\rm fin}}$.
According to the concatenation results, the current geometric center of each class is updated as
\begin{align}
\textbf{G}_{\hat{k}}^l={\rm exp}\left[\frac{1}{l}\sum_{i=1}^l{\rm Log}(\hat{\textbf{R}}_j^i)\right]
\end{align}
with $\hat{k}\in[1,\hat{K}_a], \hat{\textbf{R}}_j^i=\hat{\gamma}_{j,i}\textbf{I},j\in\hat{\mathcal{L}}_i\cap\mathcal{C}_{\hat{k}}$.
At this stage, a portion of original messages from each active UE can be obtained through unmapping.
Subsequently, codeword detection and concatenation of the next sub-slot continue until the complete messages of all sub-slots are recovered.

In summary, the proposed integrated decoding algorithm is described as Algorithm 1 and the flow chart is illustrated in Fig. \ref{Fig1}.

\section{Performance Analysis}\label{sec:ana}
In this section, to evaluate the performance of the proposed integrated decoding-based massive uncoupled unsourced random access scheme, the computational complexity and convergence rate of decoding algorithm are analyzed.

\subsection{Computational Complexity}
First, the computational complexity of the proposed integrated decoding algorithm for UURA is analyzed. One of the main computational costs is the calculation of the sample covariance matrix $\hat{\boldsymbol{\Psi}}_l^Y$, which is given by $\mathcal{O}(n_0^2M)$, but it is performed only once in each sub-slot before the start of the proximal iteration. In each iteration of proximal gradient descent, the computational complexity of the proposed decoding algorithm is dominated by three factors. Firstly, the multiplications of matrices and vectors in the calculation of gradient $\nabla f(\boldsymbol{\gamma}_l)$ have a complexity of $\mathcal{O}(n_0^2)$. Secondly, the multiplications of column vectors and row vectors in the update of covariance matrix have a complexity of $\mathcal{O}(n_0^2)$. Lastly, the calculations required by active codewords in each sub-slot to determine the class index has a complexity of $\mathcal{O}(K_a^2)$.
Therefore, the total complexity of the proposed decoding algorithm is roughly given by $\mathcal{O}(n_0^2ML+T_{\rm iter}(L-1)(n_0^2+K_a^2))$. Note that the computational cost mainly relies on the codeword length $n_0$, the number of active UEs $K_a$ and the maximum number of iteration $T_{\rm iter}$, and it does not depend on the number of potential UEs. Additionally, since the number of computations for the sample covariance matrix $\hat{\boldsymbol{\Psi}}_l^Y$ is independent of the maximum number of iteration, the increment of the number of BS antennas does not significantly increase the algorithm's complexity.

For comparison, we also analyze the computational complexity of two baselines, containing a coupled URA (CURA) scheme and a separate decoding-based UURA scheme (UURA-SD) \cite{UN5,UN11}. The settings and simulations of them are provided in the next section. From Table \uppercase\expandafter{\romannumeral1}, it can be seen that the complexity of the proposed integrated decoding-based UURA scheme (UURA-ID) is superior to the other two schemes. Herein, $\mathcal{O}_{\rm Tree}(Ka,L,a_l)=K_a(L-1)+K_a\sum\nolimits_{n=2}^{L-1}\sum\limits_{m=2}^n K_a^{n-m}(K_a-1)\prod\nolimits_{l=m}^n(2^{-a_l})$ is the complexity of tree decoding in CURA scheme with $a_l$ denoting the length of parity bits in sub-block $l$.
Although the complexity of the CURA scheme has a nearly linear relationship with the number of active devices $K_a$, it's essential to note that the factor $2^J$ significantly amplifies its overall complexity.

\begin{table}
\centering
\caption{The complexity comparison.}
\resizebox{\linewidth}{!}{
\begin{tabular}{cc}
\hline
Algorithms &  Complexity  \\
\hline
UURA-ID(proposed) & $\mathcal{O}(n_0^2ML+T_{\rm iter}(L-1)(n_0^2+K_a^2))$\\
UURA-SD & $\mathcal{O}(L(n_0^2M+T_{\rm iter}n_0^2+MK_a^3))$\\
CURA & $\mathcal{O}(2^Jn_0LM^2)+\mathcal{O}_{\rm Tree}(Ka,L,a_l)$\\
\hline
\end{tabular}}
\end{table}

\subsection{Convergence Rate}
The convergence behaviour and rate of the proximal iteration involved in the proposed integrated decoding algorithm are given below.

Recall that the objective function $\min_{\boldsymbol{\gamma}_l}\xi(\boldsymbol{\gamma}_l)=f(\boldsymbol{\gamma}_l)+\alpha g(\boldsymbol{\gamma}_l) +\beta\phi(\boldsymbol{\gamma}_l)$ has the following iterative form
\begin{align}
\boldsymbol{\gamma}_l^{t+1}&=\boldsymbol{\gamma}_l^t-\eta_l^t\nabla f(\boldsymbol{\gamma}_l^t)-\alpha\eta_l^t\partial g(\boldsymbol{\gamma}_l^{t+1})-\beta\eta_l^t\partial\phi(\boldsymbol{\gamma}_l^{t+1})\nonumber\\
&={\rm prox}_{\eta_l^t(\alpha g + \beta\phi)}\left(\boldsymbol{\gamma}_l^t-\eta_l^t\nabla f(\boldsymbol{\gamma}_l^t)\right).
\end{align}
After Douglas-Rachford splitting, the equivalent form can be give by
\begin{align}\label{DRiter}
\bm{z}^t&={\rm prox}_{\eta_l^t(\alpha g)}\left(\boldsymbol{\gamma}_l^t-\eta_l^t\nabla f(\boldsymbol{\gamma}_l^t)-\eta_l^t\beta\bm{u}^t\right),\nonumber\\
\boldsymbol{\gamma}_l^{t+1}&={\rm prox}_{\eta_l^t(\beta\phi)}\left(\bm{z}^t+\eta_l^t\beta\bm{u}^t\right),\nonumber\\
\bm{u}^{t+1}&=\bm{u}^t+\frac{1}{\eta_l^t\beta}(\bm{z}^t-\boldsymbol{\gamma}_l^{t+1}).
\end{align}

For convenience of subsequent analysis, the following lemmas are given first.
\begin{lemma}
When the differentiable function $f(\bm{x})$ has a Lipschitz-continuous gradient with constant $C_f$, i.e., it is satisfied that $\|\nabla f(\bm{x})-\nabla f(\bm{y})\|\leq C_f\|\bm{x}-\bm{y}\|,\forall \bm{x},\bm{y}$, $f(\bm{x})$ has a quadratic upper bound:
\begin{align}
f(\bm{y})\leq f(\bm{x})+\nabla f(\bm{x})^T(\bm{y}-\bm{x})+\frac{C_f}{2}\|\bm{y}-\bm{x}\|^2.
\end{align}
\end{lemma}

\begin{IEEEproof}
For $\forall \bm{x},\bm{y}$, construct a scalar function
\begin{align}
h(t)=f(\bm{x}+t(\bm{y}-\bm{x})),t\in[0,1],
\end{align}
its derivative can be expressed as
\begin{align}
h'(t)=\nabla f(\bm{x}+t(\bm{y}-\bm{x}))^T(\bm{y}-\bm{x}).
\end{align}
Due to $h(0)=f(\bm{x})$ and $h(1)=f(\bm{y})$, we have
\begin{align}
&\ \ f(\bm{y})-f(\bm{x})-\nabla f(\bm{x})^T(\bm{y}-\bm{x})\nonumber\\
&=h(1)-h(0)-\nabla f(\bm{x})^T(\bm{y}-\bm{x})\nonumber\\
&=\int_0^1h'(t)dt-\nabla f(\bm{x})^T(\bm{y}-\bm{x})\nonumber\\
&=\int_0^1\left(\nabla f(\bm{x}+t(\bm{y}-\bm{x}))^T(\bm{y}-\bm{x})-\nabla f(\bm{x})^T(\bm{y}-\bm{x})\right)dt\nonumber\\
&=\int_0^1\left((\nabla f(\bm{x}+t(\bm{y}-\bm{x}))-\nabla f(\bm{x}))^T(\bm{y}-\bm{x})\right)dt\nonumber\\
&\leq\int_0^1\|\nabla f(\bm{x}+t(\bm{y}-\bm{x}))-\nabla f(\bm{x})\|\|\bm{y}-\bm{x}\|dt\nonumber\\
&\leq\int_0^1tC_f\|\bm{y}-\bm{x}\|^2dt\nonumber\\
&=\frac{C_f}{2}\|\bm{y}-\bm{x}\|^2.
\end{align}
Therefore, we obtain
\begin{align}
f(\bm{y})\leq f(\bm{x})+\nabla f(\bm{x})^T(\bm{y}-\bm{x})+\frac{C_f}{2}\|\bm{y}-\bm{x}\|^2.
\end{align}
\end{IEEEproof}

\begin{lemma}
According to the first-order condition of convex optimization problem, when the global objective function converges to the minimum value, the sub-gradients of the local cost functions also approach zero. That is, $\bm{0}\in\nabla f(\boldsymbol{\gamma}_l^*)+\alpha\partial g(\boldsymbol{\gamma}_l^*)+\beta\partial\phi(\boldsymbol{\gamma}_l^*)$ if $\boldsymbol{\gamma}_l^*$ is the fixed point of objective function.
\end{lemma}

\begin{IEEEproof}
The iterative form of the proposed proximal operation after Douglas-Rachford splitting in \eqref{DRiter} can be rewritten as
\begin{align}
\boldsymbol{\gamma}_l^{t+1}&=F(\boldsymbol{\gamma}_l^t)\nonumber\\
&=\boldsymbol{\gamma}_l^t + {\rm prox}_{\eta_l^t(\beta\phi)}\left(2{\rm prox}_{\eta_l^t(\alpha g)}(\boldsymbol{\gamma}_l^t-\eta_l^t\nabla f(\boldsymbol{\gamma}_l^t))-\boldsymbol{\gamma}_l^t\right)\nonumber\\
&- {\rm prox}_{\eta_l^t(\alpha g)}\left(\boldsymbol{\gamma}_l^t-\eta_l^t\nabla f(\boldsymbol{\gamma}_l^t)\right).
\end{align}
Since $\boldsymbol{\gamma}_l^*$ is the fixed point, we have
\begin{align}
\boldsymbol{\gamma}_l^*&=F(\boldsymbol{\gamma}_l^*)\nonumber\\
&=\boldsymbol{\gamma}_l^* + {\rm prox}_{\eta_l^t(\beta\phi)}\left(2{\rm prox}_{\eta_l^t(\alpha g)}(\boldsymbol{\gamma}_l^*-\eta_l^t\nabla f(\boldsymbol{\gamma}_l^*))-\boldsymbol{\gamma}_l^*\right)\nonumber\\
&- {\rm prox}_{\eta_l^t(\alpha g)}\left(\boldsymbol{\gamma}_l^*-\eta_l^t\nabla f(\boldsymbol{\gamma}_l^*)\right),
\end{align}
i.e.,
\begin{align}
&\ \ \ \ {\rm prox}_{\eta_l^t(\beta\phi)}\left(2{\rm prox}_{\eta_l^t(\alpha g)}(\boldsymbol{\gamma}_l^*-\eta_l^t\nabla f(\boldsymbol{\gamma}_l^*))-\boldsymbol{\gamma}_l^*\right)\nonumber\\
&={\rm prox}_{\eta_l^t(\alpha g)}\left(\boldsymbol{\gamma}_l^*-\eta_l^t\nabla f(\boldsymbol{\gamma}_l^*)\right).
\end{align}
Define ${\rm prox}_{\eta_l^t(\alpha g)}\left(\boldsymbol{\gamma}_l^*-\eta_l^t\nabla f(\boldsymbol{\gamma}_l^*)\right)=\boldsymbol{\epsilon}$, according to the optimality condition of the proximal operator, we have
\begin{align}
\boldsymbol{\gamma}_l^*-\eta_l^t\nabla f(\boldsymbol{\gamma}_l^*)-\boldsymbol{\epsilon}\in\eta_l^t\alpha\partial g(\boldsymbol{\gamma}_l^*)
\end{align}
and
\begin{align}
\boldsymbol{\epsilon}-\boldsymbol{\gamma}_l^*\in\eta_l^t\beta\partial \phi(\boldsymbol{\gamma}_l^*).
\end{align}
Adding the left and right sides of the above two equations respectively, we can obtain
\begin{align}
\bm{0}\in\nabla f(\boldsymbol{\gamma}_l^*)+\alpha\partial g(\boldsymbol{\gamma}_l^*)+\beta\partial\phi(\boldsymbol{\gamma}_l^*).
\end{align}
\end{IEEEproof}

Now, we provide a theorem to indicate the convergence rate of the proposed integrated decoding algorithm.
\begin{theorem}
If the following assumptions hold,
\begin{itemize}
\item $f(\boldsymbol{\gamma}_l)$ is convex and its gradient is Lipschitz-continuous, i.e., $\|\nabla f(\boldsymbol{\gamma}_l^{t+1})-\nabla f(\boldsymbol{\gamma}_l^t)\|\leq C_f\|\boldsymbol{\gamma}_l^{t+1}-\boldsymbol{\gamma}_l^t\|$;
\item $g(\boldsymbol{\gamma}_l)$ and $\phi(\boldsymbol{\gamma}_l)$ are convex and their proximal operators ${\rm prox}_g,{\rm prox}_\phi$ exist;
\item the objective function $\xi(\boldsymbol{\gamma}_l)=f(\boldsymbol{\gamma}_l)+\alpha g(\boldsymbol{\gamma}_l) +\beta\phi(\boldsymbol{\gamma}_l)$ has a minimum value $\xi(\boldsymbol{\gamma}_l)^*$, which is taken at the fixed point $\boldsymbol{\gamma}_l^*$. That is, $\xi(\boldsymbol{\gamma}_l)^*=\xi(\boldsymbol{\gamma}_l^*)$,
\end{itemize}
and the iteration step size $\eta_l^t$ meets the condition that $\eta_l^t=\eta_l\in(0,1/C_f]$, we have
\begin{align}
\xi(\boldsymbol{\gamma}_l^t)-\xi(\boldsymbol{\gamma}_l^*)\leq\frac{1}{2t\eta_l}\|\boldsymbol{\gamma}_l^0-\boldsymbol{\gamma}_l^*\|^2.
\end{align}
This means that when the value of step size $\eta_l$ satisfies the interval condition constrained by Lipschitz-continuous constant $C_f$, the objective function of the proposed algorithm converges to $\xi(\boldsymbol{\gamma}_l^*)$ at the rate of $\mathcal{O}(1/t)$ and the step size affects the convergence rate.
\end{theorem}

\begin{IEEEproof}
Based on the above assumptions and Lemmas 1-2, define the generalized gradient function as
\begin{align}
G_{\eta_l}(\boldsymbol{\gamma}_l^t)=\nabla f(\boldsymbol{\gamma}_l^t)+\alpha\partial g(\boldsymbol{\gamma}_l^t)+\beta\partial\phi(\boldsymbol{\gamma}_l^t),
\end{align}
which indicates the equivalent search direction and can be expressed equivalently as follows
\begin{align}
G_{\eta_l^t}(\boldsymbol{\gamma}_l^t)=\frac{1}{\eta_l}\left[\boldsymbol{\gamma}_l-{\rm prox}_{\eta_l(\alpha g+\beta\phi)}(\boldsymbol{\gamma}_l-\eta_l\nabla f(\boldsymbol{\gamma}_l))\right].
\end{align}
Thereby, the proximal gradient iterations can be rewritten as
\begin{align}
\boldsymbol{\gamma}_l^{t+1}&=\boldsymbol{\gamma}_l^t-\eta_lG_{\eta_l}(\boldsymbol{\gamma}_l^t)\nonumber\\
&={\rm prox}_{\eta_l(\alpha g+\beta\phi)}(\boldsymbol{\gamma}_l-\eta_l\nabla f(\boldsymbol{\gamma}_l)).
\end{align}
In this case, let $\bm{y}=\boldsymbol{\gamma}_l-\eta_lG_{\eta_l}(\boldsymbol{\gamma}_l)$ and $\bm{x}=\boldsymbol{\gamma}_l$ in Lemma 1, we have
\begin{align}
f\left(\boldsymbol{\gamma}_l-\eta_lG_{\eta_l}(\boldsymbol{\gamma}_l)\right)&\leq f(\boldsymbol{\gamma}_l)-\eta_l\nabla f(\boldsymbol{\gamma}_l)^TG_{\eta_l}(\boldsymbol{\gamma}_l)\nonumber\\
&+\frac{C_f\eta_l^2}{2}\|G_{\eta_l}(\boldsymbol{\gamma}_l)\|^2.
\end{align}
Due to $\eta_l^t=\eta_l\in(0,1/C_f]$, we can obtain
\begin{align}\label{pr1}
f\left(\boldsymbol{\gamma}_l-\eta_lG_{\eta_l}(\boldsymbol{\gamma}_l)\right)&\leq f(\boldsymbol{\gamma}_l)-\eta_l\nabla f(\boldsymbol{\gamma}_l)^TG_{\eta_l}(\boldsymbol{\gamma}_l)\nonumber\\
&+\frac{\eta_l}{2}\|G_{\eta_l}(\boldsymbol{\gamma}_l)\|^2.
\end{align}
Based on the convexity of functions $f(\boldsymbol{\gamma}_l)$ , $g(\boldsymbol{\gamma}_l)$ and $\phi(\boldsymbol{\gamma}_l)$, the following inequalities are naturally satisfied
\begin{align}
f(\bm{z})&\geq f(\boldsymbol{\gamma}_l)+\nabla f(\boldsymbol{\gamma}_l)^T(\bm{z}-\boldsymbol{\gamma}_l),\nonumber\\
g(\bm{z})&\geq g(\boldsymbol{\gamma}_l-\eta_lG_{\eta_l}(\boldsymbol{\gamma}_l))\nonumber\\
&+\partial g(\boldsymbol{\gamma}_l-\eta_lG_{\eta_l}(\boldsymbol{\gamma}_l))^T(\bm{z}-\boldsymbol{\gamma}_l+\eta_lG_{\eta_l}(\boldsymbol{\gamma}_l)),\nonumber\\
\phi(\bm{z})&\geq \phi(\boldsymbol{\gamma}_l-\eta_lG_{\eta_l}(\boldsymbol{\gamma}_l))\nonumber\\
&+\partial \phi(\boldsymbol{\gamma}_l-\eta_lG_{\eta_l}(\boldsymbol{\gamma}_l))^T(\bm{z}-\boldsymbol{\gamma}_l+\eta_lG_{\eta_l}(\boldsymbol{\gamma}_l)).
\end{align}
Thus, we have
\begin{align}\label{pr2}
f(\boldsymbol{\gamma}_l)\leq f(\bm{z})-\nabla f(\boldsymbol{\gamma}_l)^T(\bm{z}-\boldsymbol{\gamma}_l)
\end{align}
and
\begin{align}\label{pr3}
&\quad g(\boldsymbol{\gamma}_l-\eta_lG_{\eta_l}(\boldsymbol{\gamma}_l)) +\phi(\boldsymbol{\gamma}_l-\eta_lG_{\eta_l}(\boldsymbol{\gamma}_l))\nonumber\\
&\leq g(\bm{z})+\phi(\bm{z})\nonumber\\
&\quad\quad -(G_{\eta_l}(\boldsymbol{\gamma}_l)-\nabla f(\boldsymbol{\gamma}_l))^T(\bm{z}-\boldsymbol{\gamma}_l+\eta_lG_{\eta_l}(\boldsymbol{\gamma}_l)).
\end{align}
Then, by adding the formulas \eqref{pr1}, \eqref{pr2} and \eqref{pr3}, the following result can be obtained
\begin{align}
\xi(\boldsymbol{\gamma}_l-\eta_lG_{\eta_l}(\boldsymbol{\gamma}_l))&\leq\xi(\bm{z}) +G_{\eta_l}(\boldsymbol{\gamma}_l)^T(\boldsymbol{\gamma}_l-\bm{z})\nonumber\\
&-\frac{\eta_l}{2}\|G_{\eta_l}(\boldsymbol{\gamma}_l)\|^2.
\end{align}
It is observed that

$\cdot$ if $\bm{z}=\boldsymbol{\gamma}_l$, we have
\begin{align}
\xi(\boldsymbol{\gamma}_l-\eta_lG_{\eta_l}(\boldsymbol{\gamma}_l))\leq\xi(\boldsymbol{\gamma}_l)-\frac{\eta_l}{2}\|G_{\eta_l}(\boldsymbol{\gamma}_l)\|^2.
\end{align}
That is, the objective function $\xi(\boldsymbol{\gamma}_l)$ is non-increasing;

$\cdot$ if $\bm{z}=\boldsymbol{\gamma}_l^*$, define $\boldsymbol{\gamma}_l^i=\boldsymbol{\gamma}_l^{i-1}-\eta_lG_{\eta_l}(\boldsymbol{\gamma}_l^{i-1}),i=1,...,t$ and we have
\begin{align}
&\quad\xi(\boldsymbol{\gamma}_l^i)-\xi(\boldsymbol{\gamma}_l^*)\nonumber\\
&=\xi(\boldsymbol{\gamma}_l^{i-1}-\eta_lG_{\eta_l}(\boldsymbol{\gamma}_l^{i-1}))-\xi(\boldsymbol{\gamma}_l^*)\nonumber\\
&\leq G_{\eta_l}(\boldsymbol{\gamma}_l^{i-1})^T(\boldsymbol{\gamma}_l^{i-1}-\boldsymbol{\gamma}_l^*)-\frac{\eta_l}{2}\|G_{\eta_l}(\boldsymbol{\gamma}_l^{i-1})\|^2\nonumber\\
&=\frac{1}{2\eta_l}\left(\|\boldsymbol{\gamma}_l^{i-1}-\boldsymbol{\gamma}_l^*\|^2-\|\boldsymbol{\gamma}_l^{i-1}-\boldsymbol{\gamma}_l^*-\eta_lG_{\eta_l}(\boldsymbol{\gamma}_l^{i-1})\|^2\right)\nonumber\\
&=\frac{1}{2\eta_l}\left(\|\boldsymbol{\gamma}_l^{i-1}-\boldsymbol{\gamma}_l^*\|^2-\|\boldsymbol{\gamma}_l^i-\boldsymbol{\gamma}_l^*\|^2\right).
\end{align}
The summation of $i=1,...,t$ in the above equation can be expressed as
\begin{align}
&\ \ \sum_{i=1}^t\left[\xi(\boldsymbol{\gamma}_l^i)-\xi(\boldsymbol{\gamma}_l^*)\right]\nonumber\\
&\leq\frac{1}{2\eta_l}\sum_{i=1}^t\left(\|\boldsymbol{\gamma}_l^{i-1}-\boldsymbol{\gamma}_l^*\|^2-\|\boldsymbol{\gamma}_l^i-\boldsymbol{\gamma}_l^*\|^2\right) \nonumber\\
&=\frac{1}{2\eta_l}\left(\|\boldsymbol{\gamma}_l^0-\boldsymbol{\gamma}_l^*\|^2-\|\boldsymbol{\gamma}_l^t-\boldsymbol{\gamma}_l^*\|^2\right) \nonumber\\
&\leq\frac{1}{2\eta_l}\|\boldsymbol{\gamma}_l^0-\boldsymbol{\gamma}_l^*\|^2.
\end{align}
According to the fact that $\xi(\boldsymbol{\gamma}_l)$ is non-increasing mentioned before, we have
\begin{align}
&\ \ \xi(\boldsymbol{\gamma}_l^t)-\xi(\boldsymbol{\gamma}_l^*)\nonumber\\
&\leq\frac{1}{t}\sum_{i=1}^t\left[\xi(\boldsymbol{\gamma}_l^i)-\xi(\boldsymbol{\gamma}_l^*)\right]\nonumber\\
&\leq\frac{1}{2t\eta_l}\|\boldsymbol{\gamma}_l^0-\boldsymbol{\gamma}_l^*\|^2.
\end{align}
The proof is completed.
\end{IEEEproof}

\section{Numerical Results}\label{sec:sim}
In this part, we conduct extensive simulations to evaluate the effectiveness of the proposed MIG-aided integrated decoding-based uncoupled unsourced random access scheme in 6G massive communications. Considering the sporadic traffic of IoT services, it is assumed that the fraction of active UEs in a cell is $K_a/K_{\textrm{tot}}=10\%$. This implies that the number of potential devices supported by the network is $K_{\textmd{tot}}=10K_a$. Unless extra specification, the main simulation parameters are set as: the number of active UEs $K_a=50$, the number of BS antennas $M=32$, the length of binary message $b=96$ bits and sub-block $J=12$ bits. The message is divided into $L=8$ sub-blocks with $n_0=100$ symbol transmissions each. The iteration step size is fixed as $\eta_l^t=\eta=0.03$.
The large-scale fading coefficient of the uplink channel for UE $k$ is $\tilde{g}_k[\textmd{dB}]=-128.1-37.6\log_{10}(d_k)$\cite{GF3}, where $d_k$ is the distance between UE $k$ and the BS and it is randomly distributed in $[0,0.5]$km.
The receive signal-to-noise ratio (SNR) of active UE is set as $10\log_{10}(P_k\tilde{g}_k/\sigma^2)=10\ {\rm dB}$.
Additionally, the codebook matrix $\textbf{C}$ is generated by $n_0$ randomly selected rows of an $2^J$-point DFT matrix with columns of unit norm. The structure of the codebook does not affect the performance of the proposed algorithm.
For UURA system, the dedicated error probability of decoding is adopted as the performance indicator to measure the average fraction of mis-decoded messages over the number of active devices. It is the sum of the per-user probability of misdetection $P_{\rm md}=\tfrac{1}{K_a}\sum\nolimits_{k\in \mathcal{K}_a}\mathbb{P}(\bm{m}_k\notin\hat{\mathcal{L}})$ and per-user probability of false alarm $P_{\rm fa}=|\hat{\mathcal{L}}\setminus\{\bm{m}_k\in \hat{\mathcal{L}}\}| / |\hat{\mathcal{L}}|$, where $\hat{\mathcal{L}}$ denotes the collection of estimated messages $\hat{\bm{m}}_k$.


\begin{figure}[h] \centering
\includegraphics [width=0.41\textwidth] {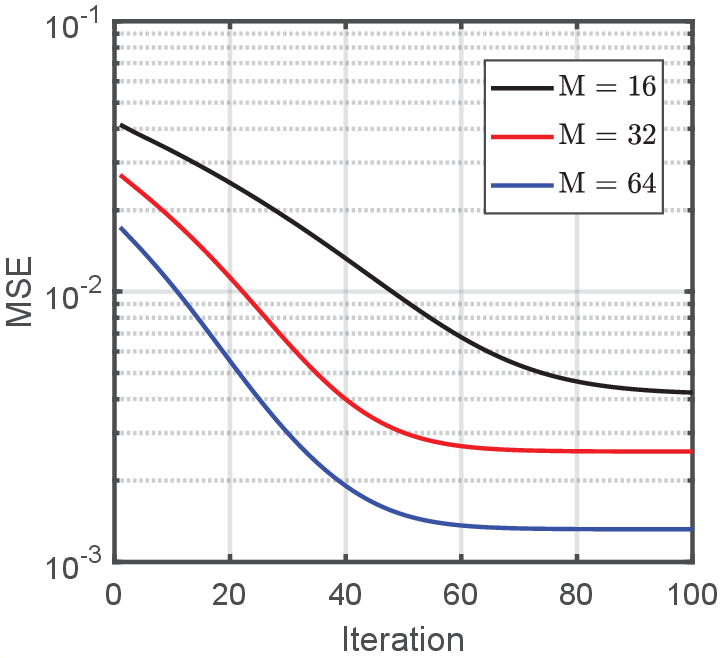}
\caption {The convergence behaviour of the proposed integrated decoding algorithm for different numbers of BS antennas.}
\label{Fig2}
\end{figure}
We first observe the convergence behaviour of the proposed integrated decoding algorithm with different numbers of BS antennas. As shown in Fig. \ref{Fig2}, the mean squared error (MSE) of the codeword activity vector $\boldsymbol{\gamma}_l^t$, i.e., $\textmd{MSE}=\tfrac{1}{2^J}\|\boldsymbol{\gamma}_l^t-\boldsymbol{\gamma}_l\|_2^2$, converges within 60, 65 and 94 iterations with precision $\varpi_2=1e-5$ for $M=64,\ 32$ and $16$, respectively. When the number of BS antennas varies, the convergence rate and fixed point of the proposed algorithm also differ. Specifically, a larger number of antennas results in a faster convergence rate and a smaller fixed point of MSE. These facts imply the feasibility of the proposed integrated decoding algorithm.
Note that the design of the decoding algorithm aims to reduce the processing delay of sub-messages at the receiver and cannot change the transmission delay of the messages over the air interface.

\begin{figure}[h] \centering
\includegraphics [width=0.41\textwidth] {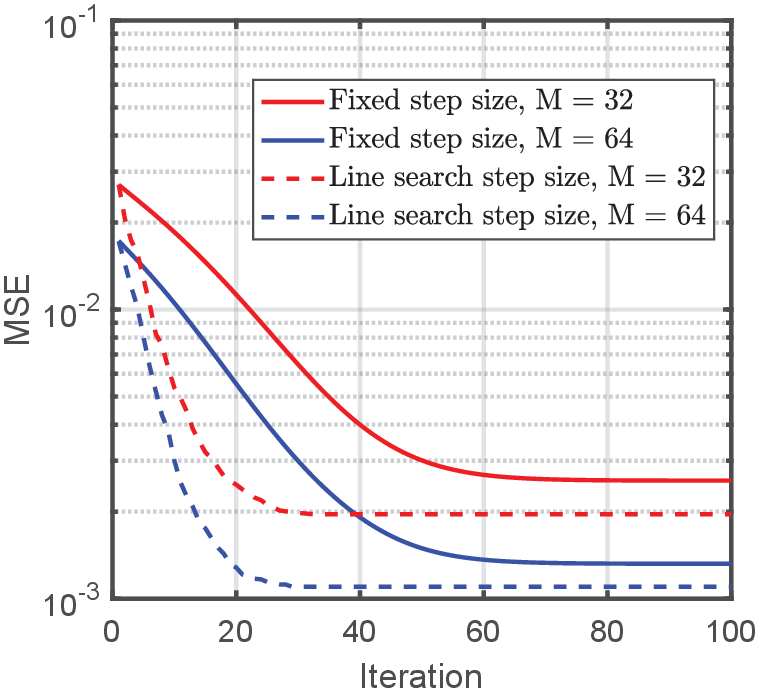}
\caption {The convergence rate of the proposed integrated decoding algorithm for different step size settings.}
\label{Fig3}
\end{figure}
Fig. \ref{Fig3} validates the influence of the step size on the convergence rate of the proposed integrated decoding algorithm. In this evaluation, we conduct the proposed decoding algorithm with a fixed step size of $\eta_l^t=\eta=0.03$ and another dynamic line search step size given by $\eta_l^{t+1}=0.8\eta_l^t$ respectively. The results demonstrate that the algorithm with the line search step size converges at a faster rate compared to the one with a fixed step size. This improvement can be attributed to the dynamic search, which enables the proposed algorithm to select a more appropriate step size in each iteration, thereby reducing the required total number of iterations. Consequently, the computational cost of the proposed algorithm is further reduced as well.

\begin{figure}[h] \flushright
\includegraphics [width=0.45\textwidth] {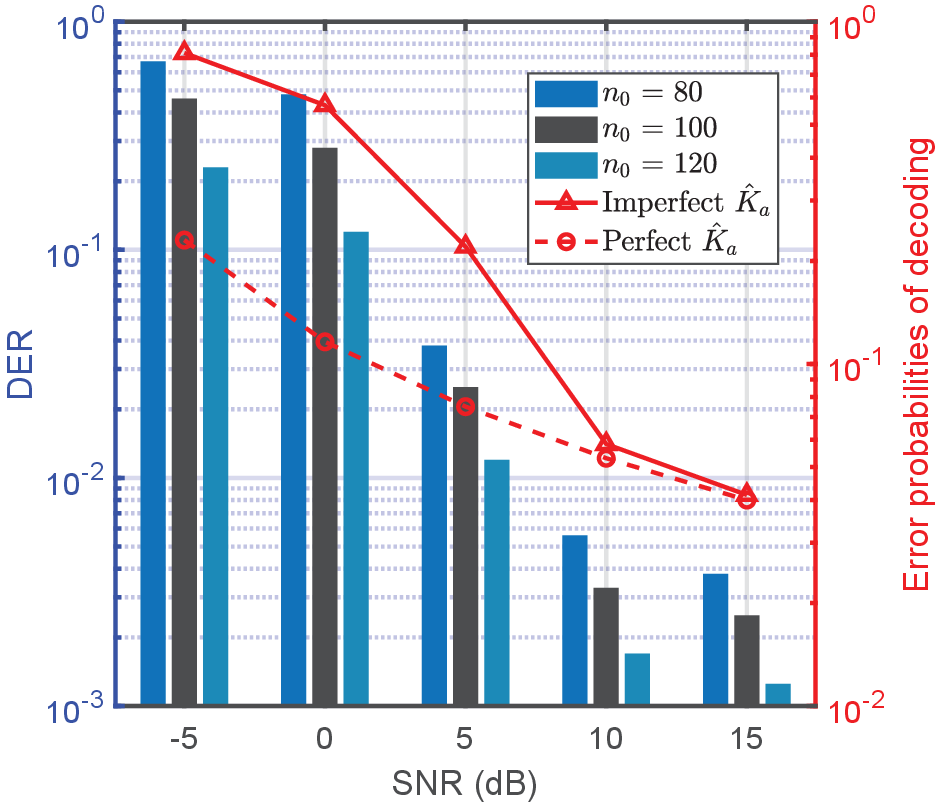}
\caption {The DER versus SNR for different lengths of codeword and error probabilities of decoding versus SNR for perfect and imperfect $\hat{K}_a$.}
\label{Fig8}
\end{figure}
As previously mentioned, the accuracy of estimated $K_a$ obtained through the decoding in the first sub-slot is crucial for the integrated detection and stitching in the subsequent sub-slots.
In Fig. \ref{Fig8}, we examine the effects of SNR and codeword length on the accuracy of $\hat{K}_a$. The detection error rate (DER) in the first sub-slot is adopted as a performance measure, which accounts for both missed detection and false alarm probabilities, i.e., when the obtained $\hat{K}_a$ is smaller or larger than the true $K_a$. Intuitively, increasing SNR and codeword length improve the DER, reaching values of $10^{-3}$ and below when SNR $\geq10$ dB.
Additionally, we investigate the impact of an inaccurate $\hat{K}_a$ on the error probabilities of the overall decoding algorithm. The red dashed line represents the situation where a high SNR is used in the first sub-slot (yielding an almost perfect $\hat{K}_a$), while the SNR shown on the x-axis is used in the other sub-slots. This line demonstrates the influence of only the SNR on the overall decoding algorithm. On the other hand, the red solid line indicates the situation where the SNR shown on the x-axis is used in all sub-slots (maybe resulting in an imperfect $\hat{K}_a$). This line reflects the combined influence of both the estimation error of $\hat{K}_a$ and the SNR on the overall decoding algorithm.
In the low SNR region, the gap between the solid line and the dashed line is significant due to the high DER, which is caused by the error in estimating $\hat{K}_a$. In contrast, in the high SNR region, the two lines almost coincide, indicating that the estimation error of $\hat{K}_a$ is small enough to be negligible. Therefore, it is advisable to assign a higher SNR to the transmitted codewords of active UEs in the first sub-slot to ensure an accurate estimation of $\hat{K}_a$. In subsequent simulations, we consider this approach to examine the individual influence of each parameter on the decoding performance.

\begin{figure}[h] \centering
\includegraphics [width=0.405\textwidth] {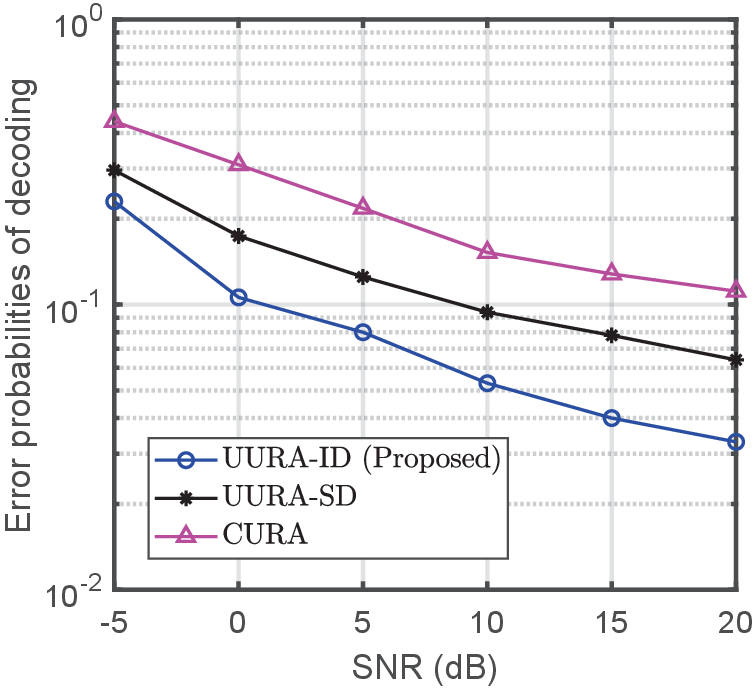}
\caption {The error probabilities of decoding versus SNR for different schemes.}
\label{Fig4}
\end{figure}
Fig. \ref{Fig4} presents the error probabilities of decoding versus SNR for various URA schemes. For comparison, we conduct the following baselines, each with parameter settings consistent with this paper unless specifically stated.\\
(i) Coupled URA (CURA) scheme: This scheme employs an energy detector for codeword detection and utilizes the tree code to couple the adjacent sub-messages by introducing additional parity check bits so that the codeword concatenation can be realized through tree decoding. To maintain the same total length of message, the message is divided into $L=32$ sub-blocks of length $J=12$ bits each. The lengths of parity check bits in $L$ sub-blocks are given by $\{0,9,...,9,12,12,12\}$.\\
(ii) Separate decoding-based UURA (UURA-SD) scheme: In this scheme, ML-based codeword detection is first implemented, followed by the codeword stitching using the K-means clustering method when all the complete messages are available.\\
It is observed that the decoding performance of all three schemes improves with increasing SNR of the active UEs. Notably, the proposed integrated decoding-based UURA (UURA-ID) scheme consistently outperforms the baselines across different SNR values. Hence, the proposed scheme is capable of providing an efficient random access solution for massive communication networks.

\begin{figure}[h] \centering
\includegraphics [width=0.41\textwidth] {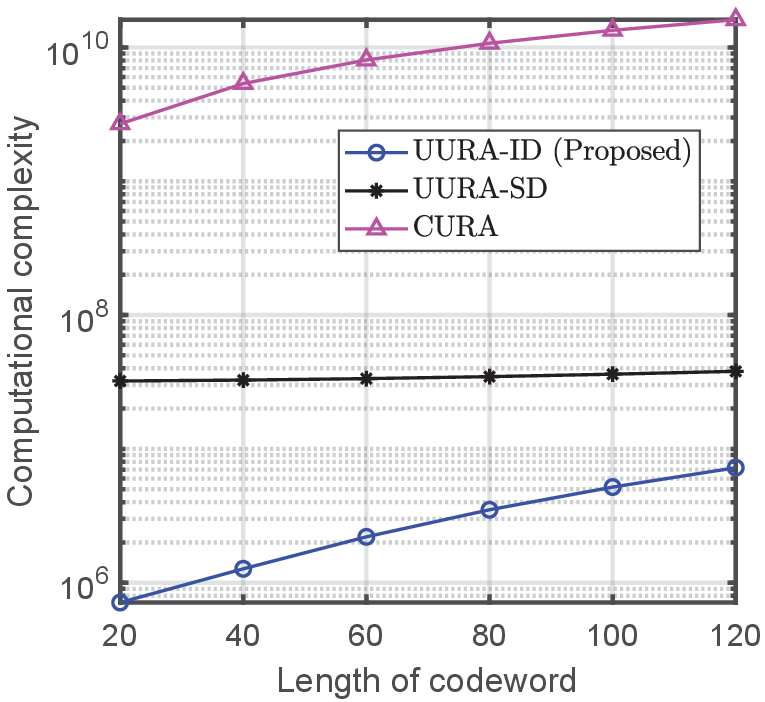}
\caption {The computational complexity versus the length of codeword $n_0$ for different schemes.}
\label{Fig5}
\end{figure}
Then, we show the computational efficiency of the proposed UURA-ID scheme compared with the two baselines. In Fig. \ref{Fig5}, it is seen that the computational costs, measured in terms of the number of multiplications required, increase as the codeword length $n_0$ increases.
Note that the proposed UURA-ID scheme exhibits the lowest number of multiplications among the three schemes. The gap in computational cost between UURA-ID and UURA-SD diminishes, while the gap between UURA-SD and CURA widens. These findings indicate that the proposed UURA-ID scheme achieves higher computational efficiency in situations with limited wireless resources.

\begin{figure}[h] \centering
\includegraphics [width=0.41\textwidth] {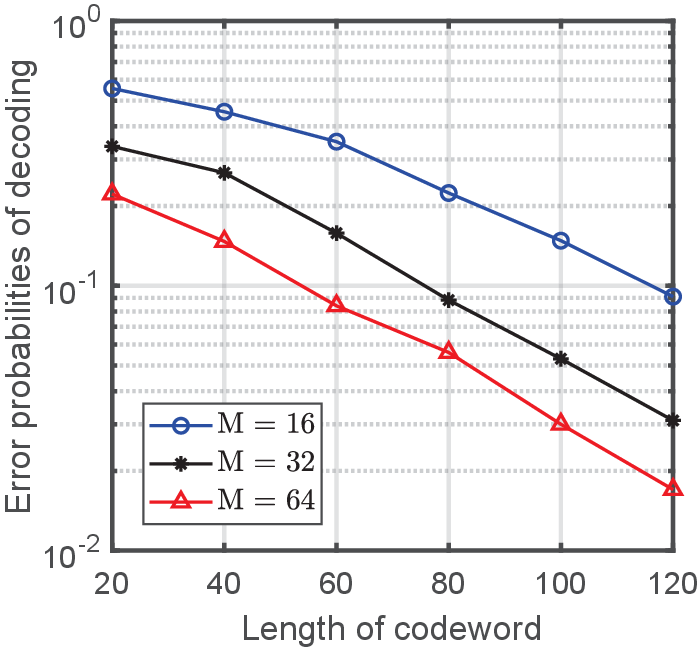}
\caption {The error probabilities of decoding versus the length of codeword $n_0$ for different numbers of BS antennas.}
\label{Fig6}
\end{figure}

Next, we check the impact of the codeword length $n_0$ on the decoding performance of the proposed scheme.
As depicted in Fig. \ref{Fig6}, it is evident that the performance of the proposed UURA-ID scheme improves as the codeword length increases across all cases. However, larger values of $n_0$ not only reduce the error probabilities of decoding but also lead to higher computational complexity. Therefore, a trade-off between decoding performance and implementation cost must be considered.
Furthermore, the decoding performance can be further enhanced by adding additional BS antennas, thanks to the increased array gains. Consequently, it becomes feasible to reduce the consumption of wireless resources by deploying more antennas at the BS.

\begin{figure}[h] \centering
\includegraphics [width=0.405\textwidth] {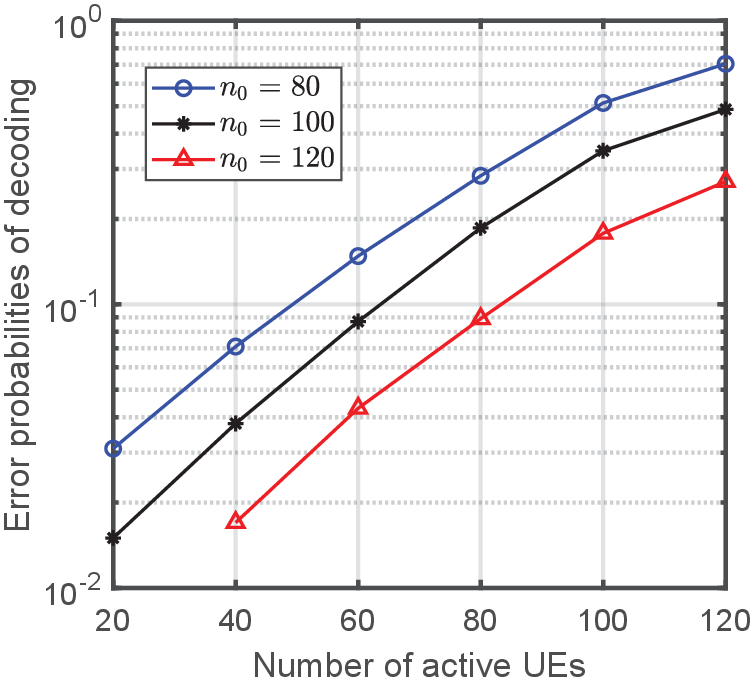}
\caption {The error probabilities of decoding versus the number of active UEs $K_a$ for different lengths of codeword $n_0$.}
\label{Fig7}
\end{figure}
Finally, Fig. \ref{Fig7} confirms the capacity of active UEs supported by the proposed UURA-ID scheme, indirectly reflecting the potential for total device capacity at a specific activity ratio.
It can be found that an increase in codeword length can effectively improve decoding performance when the number of active devices remains constant. This is because the spectral efficiency is sacrificed for performance gains. Thus, more active UEs can be accommodated by increasing the transmission resources such as the codeword length.
Moreover, the proposed UURA-ID scheme demonstrates an advantage in spectral efficiency per device due to the smaller number of required sub-blocks. This advantage allows UURA-ID to compensate for the relatively limited capacity of active devices.

In summary, the proposed MIG-aided integrated decoding algorithm has a promising potential in enhancing the performance of uncoupled unsourced random access in 6G wireless networks.

\section{Conclusion}
This paper proposed an efficient massive uncoupled unsourced random access scheme for 6G massive communications, avoiding the use of extra parity check bits. An integrated decoding algorithm for such a massive uncoupled unsourced random access scheme was designed to enable simultaneous subslot-wise codeword detection and stitching, which effectively reduced the decoding processing delay. Both theoretical analysis and numerical simulations confirmed the low complexity and good performance of the proposed scheme.

\end{document}